\documentclass[pdflatex,sn-mathphys-ay, referee, iicol]{sn-jnl}
\usepackage{graphicx}%
\usepackage{multirow}%
\usepackage{amsmath,amssymb,amsfonts}%
\usepackage{amsthm}%
\usepackage{mathrsfs}%
\usepackage[title]{appendix}%
\usepackage{xcolor}%
\usepackage{textcomp}%
\usepackage{manyfoot}%
\usepackage{booktabs}%
\usepackage{algorithm}%
\usepackage{algorithmicx}%
\usepackage{algpseudocode}%
\usepackage{listings}%
\usepackage{xcolor}
\usepackage{comment}
\usepackage{tikz}
\newcommand{\subf}[2]{%
  {\small\begin{tabular}[t]{@{}c@{}}
   \mbox{}\\[-\ht\strutbox]
   #1\\#2
   \end{tabular}}%
}
\usepackage[table]{xcolor}
\usepackage{adjustbox}
\usepackage{amsmath,amssymb}
\definecolor{lightgray}{gray}{0.95}

\theoremstyle{thmstyleone}%

\theoremstyle{thmstyletwo}%

\theoremstyle{thmstylethree}%

\begin{document}

\title[Article Title]{Constructing Large Orthogonal Minimally Aliased Response Surface Designs Through Enumeration and Combination of Weighing Designs}

\author*[1]{\fnm{Jade} \sur{LEJEUNE HERMAN}}\email{jade.lejeuneherman@kuleuven.be}
\author[1]{\fnm{Peter} \sur{GOOS}}\email{peter.goos@kuleuven.be}
\affil*[1]{\orgdiv{Mechatronics, Biostatistics and Sensors (MeBioS)}, \orgname{KU Leuven}, \orgaddress{\street{Kasteelpark Arenberg 30- bus 2456}, \city{Leuven}, \postcode{B3001}, \country{Belgium}}}

\abstract{
Advances in automation and high-throughput experimentation have enabled larger and more complex studies involving many factors and tests, creating a growing demand for computationally effective design construction methods. Efficient experimental design remains a key challenge in this context, creating a need for frameworks that can generate large experiments while preserving orthogonality and minimal aliasing. Unlike existing approaches which struggle with scalability, this work introduces an algorithmic framework for constructing large Orthogonal Minimally Aliased Response Surface (OMARS) designs by enumerating and combining weighing designs, three-level matrices with orthogonal columns and a fixed number of non-zero entries per column. Complete enumerations of weighing designs are achieved for designs with up to 24 tests, covering multiple numbers of factors and weights corresponding to two or three zeros per factor. In addition, a validated partial enumeration procedure and a combination method extend the catalog to substantially larger designs. The combination method enables the construction of OMARS designs for any test size that is a multiple of selected base sizes. This paper thus provides the methodology for generating large catalogs of high-quality OMARS designs, well-suited for high-dimensional screening and response-surface modelling in complex industrial and scientific experiments.
}




\maketitle
\newpage
\section{Introduction}\label{sec-intro}

In recent years, experimentation has entered a new stage. Traditionally, experiments were costly, time-consuming and heavily dependent on human effort. Since the early 1900s, Design of Experiments has been providing a systematic framework to make experimentation more efficient, enabling researchers to extract maximum information from a limited number of tests.
With the advent of automation, new opportunities have emerged, making it possible to conduct certain experiments on a much larger scale with more than just a handful of factors. To fully exploit this potential, experimental designs must continue to evolve and scale efficiently while preserving desirable statistical properties such as orthogonality. Whereas smaller studies necessarily focus on estimating main effects, larger studies also make it possible to explore second-order effects, including interactions and quadratic terms, which are essential for uncovering curvature and revealing the joint influence of the factors on the response.
\\ \\
One field where this shift is particularly evident is the pharmaceutical sector. 
Researchers in this field frequently rely on high-throughput screening platforms, typically organised around 96-well plates, and microarrays to conduct large numbers of tests in parallel \citep{Macarron2011}. Robotic systems take over repetitive tasks with speed and precision, enabling a higher volume of tests without a proportional increase in cost, duration or effort. When combined with carefully structured experimental designs, microarrays allow researchers to obtain deeper insights while keeping time and resource consumption under control.
\\ 
Similar opportunities for larger experiments also arise in several domains other than the pharmaceutical sector. 
In materials science, high-throughput synthesis allows dozens of compositional factors to be screened simultaneously \citep{Potyrailo2011}. In biology, pooled assays routinely involve large factor counts \citep{Replogle2022}. In manufacturing, process optimisation studies balance many tuning parameters under tight cycle times. In each of these settings, the experimenter would ideally combine screening with response surface estimation in a single experiment. Screening identifies which of many factors actually matter. Response surface estimation characterises curvature and interaction effects. The designs constructed in this paper expand the range of run sizes and factor counts at which such combined experiments become feasible. They make the single-step workflow available for problem sizes that 
existing design catalogs could not reach.
\\ \\
To make the most of these large-scale studies, the choice of experimental design remains essential. Among the various families of experimental designs, Orthogonal Minimally Aliased Response Surface (OMARS) designs stand out as particularly valuable. OMARS designs were developed to bridge the gap between definitive screening designs and traditional response surface designs (central composite and Box-Behnken designs). Their properties make them especially suitable for combining screening experimentation and response surface experimentation in a single step. This integration not only saves time and costs but also significantly accelerates research.
\\ \\
OMARS designs are three-level designs for multiple quantitative factors in which main effects are orthogonal to each other and to both two-factor interactions and quadratic effects. 
The high degree of orthogonality and limited aliasing yields statistically independent estimates of the main effects while permitting efficient exploration of second-order models. 
These properties make OMARS designs particularly effective when both curvature and interactions are present. OMARS designs also support a design-based model selection framework \citep{Hameed2023-fh}, even when the available number of runs is insufficient to estimate the full second-order response surface model.
\\ \\
The OMARS designs were introduced by \cite{Nunez_Ares2020-ig} as a generalization of definitive screening designs \citep{Jones2011-uf, Xiao2012-em, Schoen2019-te}. They proposed an enumeration procedure for OMARS designs based on integer programming. The method formulates the search for eligible designs as an integer feasibility problem, using binary decision variables to indicate whether potential design points are included and a system of linear equations to enforce the defining properties of OMARS designs. In the enumeration procedure, isomorphic designs (i.e. designs that can be generated from one another by permutations of factors, permutations of tests or sign changes of columns) are systematically excluded. 
This iterative process successfully enumerated a catalog of non-isomorphic OMARS designs for experiments involving 3 to 5 factors (with partial enumeration for 6 and 7 factors) and covered a spectrum of test sizes specific to each factor count.
\\ \\
\cite{mixedOMARS} extended the methodology to mixed-level designs, which combine three-level quantitative factors with two-level categorical factors. 
Subsequently, \cite{blockedOMARS} addressed the orthogonal blocking of both three-level and mixed-level OMARS designs.
\cite{Goos2025-gu} provided a comprehensive review of the OMARS design framework, described some constructions based on combinations of weighing matrices, distinguished between uniform- and non-uniform-precision OMARS designs and introduced the concept of strong OMARS designs. Uniform-precision OMARS designs ensure that all main effects can be estimated with equal precision. Strong OMARS designs have the additional property that interactions are orthogonal to each other and to quadratic terms, so that the main effects and the interactions can be estimated independently. 
\\ 
Inspired by the work of \cite{Xiao2012-em} for definitive screening designs, \cite{Georgiou2014-hd} showed that uniform-precision OMARS designs can be generated by concatenating a weighing matrix with its foldover. Building on this idea, the present work focuses on constructing larger uniform-precision OMARS designs through foldover of weighing designs, with a fixed number of zeros per column to preserve uniform precision. To ensure finding the best OMARS designs, an enumeration procedure for weighing designs is developed. This contrasts with \cite{Georgiou2014-hd}, who rely on readily available weighing matrices and therefore cannot guarantee that the best OMARS designs are obtained. Supporting the construction adopted in this work, \cite{Stallrich2025-vc} demonstrate that foldover constructions efficiently support screening experiments by guaranteeing zero aliasing and allowing unbiased variance estimation.
\\ \\
The remainder of this paper is structured as follows. Section 2 formulates the gap in the literature as a concrete research problem and proposes a solution based on weighing designs for the generation of larger-scale OMARS designs. Section 3 details the enumeration procedure for weighing designs and describes the corresponding algorithm. Section 4 presents the results of this enumeration, the subsequent transformation into OMARS designs and an assessment of the statistical efficiency of the resulting OMARS designs.
\section{Formal Framework}
The OMARS designs currently available are restricted in size due to the
computational constructions used to generate
them. Direct enumeration of OMARS designs becomes intractable at larger run sizes because the computational cost increases steeply with the number of runs. In contrast, enumerating weighing designs and folding them over make the search for OMARS designs feasible at substantially larger sizes for two reasons. First, a weighing design only has
half of the number of runs of an OMARS design. The enumeration of weighing designs therefore involves much smaller matrices. Second,
the defining condition for weighing designs is pairwise orthogonality of the
columns. This allows the designs to be constructed column by column, where
each candidate column only needs to be checked against the columns already
present. Weighing designs in combination with the foldover technique therefore provide a promising route for extending
OMARS constructions to larger run sizes.
In this framework, the focus is on uniform-precision OMARS designs since the structure of weighing matrices and weighing designs naturally enforces an equal occurrence of middle levels across factors. This guarantees that all main effects are estimated with the same precision.
\\ \\
A classical weighing matrix $W(m,k)$ is a square matrix of order $m$ whose entries belong to the set $\{0,\pm1\}$, in which each row and each column contain exactly $k$ non-zero entries and which satisfies the orthogonality condition $$WW'=W'W=kI_m.$$ The parameter $k$ is referred to as the weight of $W$. When $k=m-1$, in which case $W(m,m-1)$ corresponds to a conference matrix, and when $k=m$, $W(m,m)$ corresponds to a Hadamard matrix.
\\ \\
Few enumerations of weighing matrices have been carried out in the literature and only specific cases have been studied in detail. The first systematic enumeration of weighing matrices was conducted by \cite{Chan}, who classified all non-isomorphic weighing matrices of order $m \le 20$ with weight $k<5$. Later, \cite{Ohmori1992-ig} investigated the construction of weighing matrices of order $8a - 2$ and weight $4a$ for $a > 2$, providing a general solution to the intersection pattern condition and a complete classification for the special case $a = 2$, corresponding to $W(14, 8)$. In a related work, \cite{Ohmori1993-fu} provided a full classification of weighing matrices of order 13 and weight 9. Furthermore, \cite{OhmoriHiroyuki} completed the classification of weighing matrices $W(17,9)$.
\\ \\
More recently, \cite{Harada2012-gv} proposed a classification method for weighing matrices based on the classification of self-orthogonal codes. Their work extended and refined existing classifications, while providing several corrections and new insights. In particular, they revised known classifications for weighing matrices of orders up to 15 and of order 17 and also delivered a corrected classification of weighing matrices of weight 5 across all orders. Finally, \cite{Araya2025-ew} investigated a specific subclass of weighing matrices of weight 9, known as unbiased weighing matrices. 
\\ \\
The classical definition dictates that weighing matrices are square matrices. This limits the possibilities for combining them into OMARS designs while practical applications often require designs with different numbers of rows and columns. To address this issue, the concept of a weighing matrix is extended to rectangular weighing designs where the number of columns may be smaller than the number of rows. The present paper therefore generalizes \cite{Georgiou2014-hd} in two ways. First, it considers rectangular weighing designs instead of square weighing matrices. Second, whereas \cite{Georgiou2014-hd} rely on individual weighing matrices available in the literature, the present paper enumerates all non-isomorphic weighing designs within the computational limits, so that the best design can be selected for each size. It also generalizes \cite{Schoen2019-te}, who focused on conference designs, by addressing the more general case of weighing designs that include more zeros per column.
\\ \\
A weighing design $W(m,z,k)$ is defined as a matrix of size $m \times z$ with entries from $\{0,\pm1\}$, in which each column contains exactly $k$ non-zero entries and each row contains at most $m-k$ zeros. An obvious special case arises when $m=z$, in which case $W(m,z,k)$ is a weighing matrix. The columns are required to be mutually orthogonal, which is expressed by the condition $W'W=kI_z$. This condition is used rather than $WW'$ because the primary interest in DOE is column orthogonality. For square matrices ($m=z$), column orthogonality automatically implies row orthogonality so checking $W'W$ suffices. In rectangular designs with $z < m$, column orthogonality does not guarantee row orthogonality.
\\ \\
The distribution of the zeros in a weighing design has a major impact on its usefulness as an experimental design for screening and response surface experimentation. First, there should not be too many zeros, since zeros have a negative impact on the precision of main-effect estimates obtained from an OMARS design and the estimability of interaction effects. Second, for the estimation of individual quadratic effects, it is important that zeros are distributed across multiple rows of the weighing design.
\\ \\
To avoid duplicate representations arising from row sign changes or column permutations and to ensure that each weighing design is represented in a unique canonical form, the first column of the design is fixed. The first $m-k$ entries are set to zero and the last $k$ entries are set to one. Similarly, the first row is standardized so that its entries can only take the values 0 or 1.
\\ \\
One aim of this study is to generate all non-isomorphic weighing designs for each combination of the parameters $m,z$ and $k$. Two designs are considered isomorphic and belong to the same isomorphism class if one can be transformed into the other through a sequence of row permutations, column permutations or sign changes applied to rows or columns. Any two isomorphic designs share the same statistical properties and efficiency, so retaining both is not useful from a statistical viewpoint. Therefore, it is sufficient to enumerate one representative design from each isomorphism class. Row sign changes are included in the definition of isomorphism because the ultimate aim of this paper is to construct OMARS designs by combining a weighing design with its foldover: $$D=\begin{bmatrix}
    \phantom{-}W \\-W
\end{bmatrix}.$$
In such a construction, each row and its negative counterpart already appear in the matrix. Consequently, any row sign change applied to $W$ yields the same OMARS design $D$, making the two representations indistinguishable in the OMARS design context.
\\ \\
To motivate the construction of OMARS designs from weighing designs, it is useful to show that combining a weighing design with its foldover naturally yields an OMARS design. A formal proof of this property is provided in Appendix \ref{app:apA}.
\\ \\
Before the development of general OMARS designs, 
one special case had already been introduced by \cite{Jones2011-uf}: the definitive screening designs (DSDs). Later, \cite{Xiao2012-em} presented an efficient construction method for DSDs. By concatenating a conference matrix with its foldover to generate a DSD, their method provides the foundation for the present work. \cite{Phoa2015-xh} analyzed the structure of DSDs and proposed a systematic, theory-driven construction method which applies universally to matrices of any order and is computationally efficient. 
\\ \\
The contribution of this paper is therefore threefold:
\begin{itemize}
\item it develops an enumeration framework for weighing designs that extends beyond the classical square case; 
\item it uses this framework to construct and explore new uniform-precision OMARS designs suitable for modern large-scale experimentation;
\item it produces a catalog of OMARS designs as a practical deliverable 
suitable for direct use by practitioners.
\end{itemize}
For this purpose, a dedicated enumeration algorithm has been developed to generate weighing designs, identify unique representatives and provide the necessary input for the constructions and analyses presented in Sections 3 and 4.
\section{The enumeration procedure}
The enumeration procedure for generating all weighing designs of size $m \times z$ with weight $k$ consists of two main stages. It builds the designs column by column, starting from $z=1$. For each $z\ge 2$, the procedure takes as input the collection of all previously computed weighing designs of size $m \times(z-1)$ and weight $k$, denoted by $\mathcal{W}_{m,z-1, k}$. The objective is then to generate all possible one-column extensions of each matrix in $\mathcal{W}_{m,z-1, k}$, resulting in a new collection $\mathcal{W}_{m,z, k}$.
\\ \\
In the first stage, each matrix $A \in \mathcal{W}_{m,z-1,k}$ is associated with a collection $\mathcal{S}(A)$ of binary vectors, referred to as zero patterns. Each zero-pattern vector in $\mathcal{S}(A)$ defines a valid configuration of zero positions for a new column that may be appended to $A$. If $i$ denotes the row index in a zero-pattern vector $s \in \mathcal{S}(A)$, then $s_i=0$ enforces a zero in row $i$ of the new column, whereas $s_i=1$ enforces a non-zero entry ($\pm 1$) in that position. The step of assigning $\pm 1$ values to these non-zero positions will be referred to as sign assignment. The collection $\mathcal{Z}_{m,z,k}$ consists of all $m \times z$ matrices formed by appending each of the zero-pattern vectors in $\mathcal{S}(A)$ to the matrix $A$. This collection represents all admissible zero configurations for the next column, prior to sign assignment and will be referred to as the zero-pattern-extended matrices.
\\ \\
In the second stage, each matrix $B \in \mathcal{Z}_{m,z,k}$ is processed by generating all possible relevant sign assignments in the last column. The collection $\mathcal{C}(B)$ is composed of all vectors obtained by replacing every non-zero entry in the last column of $B$ with either $+1$ or $-1$. 
For each resulting vector, a candidate matrix $M$ is formed by replacing the last column of $B$ with this vector. Each candidate matrix $M$ is tested for column-wise orthogonality and is added to the collection $\mathcal{W}_{m,z,k}$ if and only if it is orthogonal and not isomorphic to any matrix already in the collection. 
\\ \\
The two stages of the enumeration procedure, namely the generation of zero-pattern extensions and the sign assignment with isomorphism filtering, are described in detail in Sections 3.1 and 3.2. These stages are carried out for each case where $z \ge 2$. The case $z=1$ is treated separately, as it corresponds to a single, fixed column $c^{(0)}$ with 0s as the first $m-k$ entries and 1s as the remaining $k$ entries. 
This fixed initial column serves as the starting point for the subsequent enumeration procedure, ensuring a consistent construction for all weighing designs considered.
\subsection{Stage 1: Enumeration of the zero-pattern-extended matrices} 
In this stage, the enumeration is performed starting from the previously generated collection of non-isomorphic weighing designs, denoted by $\mathcal{W}_{m,z-1,k}$. 
As demonstrated in Appendix \ref{app:apB}, it is sufficient to extend only one representative from each isomorphism class of size \( m \times (z-1) \) to generate all isomorphism classes of size \( m \times z \). Using this approach, the complete enumeration of designs with $z$ columns can be carried out efficiently by extending only the designs in $\mathcal{W}_{m,z-1,k}$.
\subsubsection{Three-Step Approach}
The enumeration of admissible zero-pattern-extended matrices $\mathcal{Z}_{m,z,k}$, given a collection $\mathcal{W}_{m,z-1,k}$ of weighing designs of size $m \times (z-1)$ and weight $k$ proceeds in three steps:
\begin{enumerate}
    \item For each matrix $A \in \mathcal{W}_{m,z-1,k}$, compute the row-wise zero count vector $v^{(A)} \in \mathbb{N}^m$, where the ith entry $v_i^{(A)}$ denotes the number of zero entries in the $i$-th row of $A$. This vector summarizes the zero distribution across the rows and is used to determine which positions are available for non-zero entries in the next column.
\item This step generates all zero-pattern candidate columns $s \in S(A)$ for each matrix $A \in \mathcal{W}_{m,z-1,k}$. All binary vectors of length $m$ that contain exactly $k$ ones (and thus $m-k$ zeros) are considered. These vectors are constructed by enumerating all possible bitmasks of length $m$. A bitmask is a binary representation of an integer where each bit corresponds to a position in the vector: a bit set to 1 indicates the presence of a one in that position and a bit set to 0 indicates a zero. By iterating through all integers from $0$ to $2^m-1$, every possible binary vector of length $m$ can be represented. Among these bitmasks, only those with exactly $k$ bits set to 1 are selected. Each selected bitmask is then decoded back into a binary vector $s \in \{0,1\}^m$. 
\\ \\
Not all of the generated vectors are admissible. Two exclusion criteria are applied to filter out invalid candidates:
\begin{itemize}
    \item \textit{Exclusion criterion 1:} the last $k$ entries of $s$, referred to as the tail, are analysed. The number of zeros in this tail segment is counted and the vector $s$ is accepted only if the count is even. This parity condition ensures orthogonality with the first column $c^{(0)}$ of a weighing design.
If the number of zeros in the tail were odd, no combination of $\pm1$ entries in those positions could make the vector orthogonal to $c^{(0)}$ thus failing orthogonality. \\ \\
The remaining candidates are binary vectors of length $m$ with exactly $m-k$ zeros in total and an even number of zeros in the tail. Since $k$ is always even, the parity condition is equivalent to requiring that the number of ones in the tail, denoted $r$, is also even.
Let $C_{n,p}$ denote the binomial coefficient, i.e. the number of ways to choose $p$ elements from $n$. 
For a given $r$, the number of vectors with $r$ ones in the tail is $C_{k,r}$ and the remaining $k-r$ ones can be placed among the first $m-k$ positions in $C_{m-k,k-r}$ ways. Summing over all even $r$ gives the maximum total number of candidate vectors for a single matrix $A$: $$\sum_{\substack{r=0 \\ r \ even}}^{k}C_{k,r} \ C_{m-k,k-r}.$$
    \item \textit{Exclusion criterion 2:} using the vector $v^{(A)}$ computed in Step 1 for matrix $A$, any candidate vector $s$ having a zero in position $j$ where $v^{(A)}_j=m-k$ is excluded. This ensures that no row has more than the maximum allowed number of zeros. Formally, the set of valid candidate vectors for matrix $A$ is 
\[
\begin{aligned}
\mathcal{S}(A) = {} &
\big \{ s \in \{0,1\}^m \;|\; \\
&\sum_{i=0}^{m-1} s_i = k, \\
&\sum_{i=m-k-1}^{m-1} (1-s_i) \equiv 0 \pmod{2}, \\
& s_i = 1 \;\forall i \in I
\big \}
\end{aligned}
\]

where
\[
I = \{ i \in \{0,\ldots,m-1\} \mid v_i^{(A)} = m-k \}.
\]
\end{itemize}
\item Step 3: For each matrix $A \in \mathcal{W}_{m,z-1,k}$, every candidate vector $s \in\mathcal{S}(A)$ is appended as the $z$-th column of $A$ to form an extended matrix $B$ of size $m \times z$. All resulting matrices are collected into a new set $\mathcal{Z}_{m,z,k}$ which contains all valid column extensions at this stage. This operation can be expressed mathematically as:
$$\mathcal{Z}_{m,z,k} = \{ [A \mid c] \mid \; A \in \mathcal{W}_{m,k}^{(z-1)},\, s \in \mathcal{S}(A)\}.$$
The maximum number of matrices in this collection is
$$|\mathcal{W}_{m,z-1,k}| \cdot \sum_{\substack{r=0 \\ r \ even}}^{k}C_{k,r} \ C_{m-k,k-r}.$$ 
This maximum increases with the number of matrices in $\mathcal{W}_{m,z-1,k}$. Moreover, as the number of columns $z$ increases, the set of admissible vectors $s$ becomes smaller, since the condition that every row can contain at most $m - k$ zeros progressively reduces the pool of valid zero-pattern extensions.
\end{enumerate}
\subsubsection{Illustration of Stage 1}
To illustrate the three-step enumeration procedure for the zero-pattern-extended matrices, Figure \ref{fig:3-step} presents an example with $m=6$ rows, weight $k=4$ and $z=3$ desired columns. The process starts from the collection $\mathcal{W}_{6,2,4}$ consisting of two non-isomorphic matrices of size $6 \times 2$:
\[
\mathcal{W}_{6,2,4} =
\left\{
{\renewcommand{\arraystretch}{0.7}
\begin{bmatrix}
0 & \phantom{-}0 \\
0 & \phantom{-}0 \\
1 & \phantom{-}1 \\
1 & \phantom{-}1 \\
1 & -1 \\
1 & -1 \\
\end{bmatrix}},
\quad
{\renewcommand{\arraystretch}{0.7}
\begin{bmatrix}
0 & \phantom{-}1 \\
0 & -1 \\
1 & \phantom{-}0 \\
1 & \phantom{-}0 \\
1 & \phantom{-}1 \\
1 & -1 \\
\end{bmatrix}}
\right\}.
\]
The enumeration method iterates through each matrix in this collection, but for this illustration, attention is restricted to the first matrix only. The three steps carried out by the algorithm for the first matrix (labeled $A$) are visualized in Figure \ref{fig:3-step}. 
\begin{figure*}[h!]
\setlength{\abovecaptionskip}{1pt}
\centering
\fbox{%
\begin{minipage}{0.99\linewidth}
\centering
\begin{tikzpicture}[node distance=2cm, every node/.style={font=\small}]

\node at (0.3,4) {\textbf{Step 1}};
\node ($A$) at (-0.5,1.9) {$
{\renewcommand{\arraystretch}{0.7}{\setlength{\arraycolsep}{3pt}
\begin{bmatrix}
0 & \phantom{-}0 \\
0 & \phantom{-}0 \\
1 & \phantom{-}1 \\
1 & \phantom{-}1 \\
1 & -1 \\
1 & -1
\end{bmatrix}}}
$};
\node ($v$) at (1,1.9) {$
{\renewcommand{\arraystretch}{0.7}
\begin{bmatrix}
2 \\ 2 \\ 0 \\ 0 \\ 0 \\ 0
\end{bmatrix}}
$};
\node at (0.45,1.9) {$\rightarrow$};
\node at (-0.5, 0.5) {$A$};
\node at (1,0.5) {$v$};
\draw[dashed] (1.7,0) -- (1.7,4);
\node at (5.3,4) {\textbf{Step 2}};
\node (mat1) at (3.2,1.9) {$
{\renewcommand{\arraystretch}{0.7}
{\setlength{\arraycolsep}{3pt}
\begin{matrix}
[0 & 0 & 1 & 1 & 1 & 1]\\
[1 & 1 & 0 & 0 & 1 & 1]\\
[1 & 1 & 0 & 1 & 0 & 1]\\
[1 & 1 & 0 & 1 & 1 & 0] \\
[1 & 1 & 1 & 0 & 0 & 1] \\
[1 & 1 & 1 & 0 & 1 & 0] \\
[1 & 1 & 1 & 1 & 0 & 0]
\end{matrix}}}
$};
\node at (5.2, 1.9) {$\rightarrow$};
\node at (5.3,1.4) {filtering};
\node at (5.2, 0.9) {by $v$ };
\node (mat2) at (7.2,1.9) {$
{\renewcommand{\arraystretch}{0.7}
{\setlength{\arraycolsep}{3pt}
\begin{matrix}
[1 & 1 & 0 & 0 & 1 & 1] \\
[1 & 1 & 0 & 1 & 0 & 1] \\
[1 & 1 & 0 & 1 & 1 & 0] \\
[1 & 1 & 1 & 0 & 0 & 1] \\
[1 & 1 & 1 & 0 & 1 & 0] \\
[1 & 1 & 1 & 1 & 0 & 0]
\end{matrix}}}
$};
\draw[dashed] (8.6,0) -- (8.6,4);
\node at (11.7,4) {\textbf{Step 3}};
\node (s31) at (9.7,2.8) {$
{\renewcommand{\arraystretch}{0.6}
{\setlength{\arraycolsep}{3pt}
\begin{bmatrix}
0 & \phantom{-}0 & \phantom{-}1 \\
0 & \phantom{-}0 & \phantom{-}1\\
1 & \phantom{-}1 & \phantom{-}0\\
1 & \phantom{-}1 & \phantom{-}0\\
1 & -1 & \phantom{-}1 \\
1 & -1 & \phantom{-}1
\end{bmatrix}}}
$};
\node (s32) at (11.7,2.8) {${\renewcommand{\arraystretch}{0.6}
{\setlength{\arraycolsep}{3pt}
\begin{bmatrix}
0 & \phantom{-}0 & \phantom{-}1 \\
0 & \phantom{-}0 & \phantom{-}1\\
1 & \phantom{-}1 & \phantom{-}0 \\
1 & \phantom{-}1 & \phantom{-}1\\
1 & -1 & \phantom{-}0 \\
1 & -1 & \phantom{-}1
\end{bmatrix}}}
$};
\node (s33) at (13.7,2.8) {${\renewcommand{\arraystretch}{0.6}
{\setlength{\arraycolsep}{3pt}
\begin{bmatrix}
0 & \phantom{-}0 & \phantom{-}1 \\
0 & \phantom{-}0 & \phantom{-}1 \\
1 & \phantom{-}1 & \phantom{-}0 \\
1 & \phantom{-}1 & \phantom{-}1\\
1 & -1 & \phantom{-}1\\
1 & -1 & \phantom{-}0
\end{bmatrix}}}
$};

\node (s34) at (9.7, 1) {${\renewcommand{\arraystretch}{0.6}
{\setlength{\arraycolsep}{3pt}
\begin{bmatrix}
0 & \phantom{-}0 & \phantom{-}1 \\
0 & \phantom{-}0 & \phantom{-}1\\
1 & \phantom{-}1 & \phantom{-}1\\
1 & \phantom{-}1 & \phantom{-}0 \\
1 & -1 & \phantom{-}1 \\
1 & -1 & \phantom{-}0
\end{bmatrix}}}
$};
\node (s35) at (11.7,1) {${\renewcommand{\arraystretch}{0.6}
{\setlength{\arraycolsep}{3pt}
\begin{bmatrix}
0 & \phantom{-}0 & \phantom{-}1 \\
0 & \phantom{-}0 & \phantom{-}1 \\
1 & \phantom{-}1 & \phantom{-}1 \\
1 & \phantom{-}1 & \phantom{-}0 \\
1 & -1 & \phantom{-}0 \\
1 & -1 & \phantom{-}1
\end{bmatrix}}}
$};
\node (s36) at (13.7,1) {${\renewcommand{\arraystretch}{0.6}
{\setlength{\arraycolsep}{3pt}
\begin{bmatrix}
0 & \phantom{-}0 & \phantom{-}1 \\
0 & \phantom{-}0 & \phantom{-}1 \\
1 & \phantom{-}1 & \phantom{-}1 \\
1 & \phantom{-}1 & \phantom{-}1\\
1 & -1 & \phantom{-}0\\
1 & -1 & \phantom{-}0
\end{bmatrix}}}
$};
\end{tikzpicture}
\caption{Example of the three structured steps with $m=6$, $k=4$ and $z=3$}
\label{fig:3-step}
\end{minipage}}
\end{figure*}
\\
In the first step, the row-wise zero count vector $v \in \mathbb{N}^6$ is computed for the chosen matrix $A$. Each entry $v_i$ records the number of zeros in the $i$-th row of $A$. The top two rows of $A$ contain two zeros each, while all other rows contain none, so that $v=[2,2,0,0,0,0]^T$. This vector determines which rows are blocked from receiving additional zeros in subsequent steps.
\\ \\
The second step enumerates all binary vectors of length 6 containing exactly $k=4$ ones and $m-k=2$ zeros. These vectors represent zero-pattern candidate columns to be appended to the matrix $A$. Among them, only those with an even number of zeros in the last $k=4$ entries (the tail) are retained to allow for orthogonality with the first column of the design. This yields the seven candidate vectors displayed as a row vector in the middle panel of Figure \ref{fig:3-step}. Next, any vector that assigns a zero to a row $j$ where $v_j^{(A)}=2$ is removed, since that row has already reached the maximum number of allowed zeros. After this filtering, six valid candidates remain.
\\ \\
In the third step, each of the six remaining candidate vectors is appended as a third column to the original matrix $A$, producing six zero-pattern-extended matrices of size $6 \times 3$. These matrices form a subset of the new collection $\mathcal{Z}_{6,3,4}$, which contains all admissible extensions of matrices in $\mathcal{W}_{6,2,4}$ by one column. That subset is shown in the rightmost panel of Figure \ref{fig:3-step}. The three steps are repeated independently for the other matrix in the original collection $\mathcal{W}_{6,2,4}$. The next step is to perform the sign assignment in the final columns of the new matrices, in the second stage of the enumeration algorithm.
\subsection{Stage 2: Enumeration of the weighing designs}
The aim of this second stage is to transform the set of zero-pattern-extended matrices $\mathcal{Z}_{m,z,k}$ into the collection of the non-isomorphic weighing designs $\mathcal{W}_{m,z,k}$. The transformation is achieved by replacing each $1$ in the last column of the matrices in $\mathcal{Z}_{m,z,k}$ with either $+1$ or $-1$, while keeping the $0$ entries remain unchanged. Since all columns must be mutually orthogonal and the first column is fixed, the number of potential columns to generate can be reduced by producing only candidates that are orthogonal to the first column $c^{(0)}$ which contains $m-k$ consecutive zeros followed by $k$ consecutive ones. For each matrix $B \in \mathcal{Z}_{m,z,k}$, the last column $s$ is split in two parts: the head, corresponding to the first $m-k$ elements and the tail, corresponding to the last $k$ elements. This distinction is important because the entries in the head do not contribute to the inner product with the first column, as they correspond to positions where the first column has zeros. Indeed, if the head is denoted by $h=(h_0,\ldots,h_{m-k-1})$ and if the tail is denoted by $t=(t_0,\ldots,t_{k-1})$, the inner product between the first column $c^{(0)}$ and the last column $s$ can be written as $$c^{(0)} \cdot s=\sum_{i=0}^{m-k-1}0 \times h_i+\sum_{i=0}^{k-1}1 \times t_i=\sum_{i=0}^{k-1}t_i.$$
The orthogonality condition with the first column reduces to 
\begin{equation}
    \sum_{i=0}^{k-1}t_i= 0.
    \label{refeq2}
\end{equation}
The orthogonality condition therefore imposes a constraint solely on the tail vector $t$, while the head $h$ remains unrestricted. Depending on the distribution of zeros between the head and the tail, two distinct cases can be identified:
\begin{itemize}
\item \underline{Case 1:} All zero entries in the last column of $B$ appear in the head, which means the head consists entirely of zeros. The tail then contains only non-zero elements, each equal to either $+1$ or $-1$. According to the orthogonality condition in Equation (\ref{refeq2}), the elements of the tail must sum to zero. This implies that the tail vector must be balanced and that $k$ must be even. The number of possible tail vectors corresponds to the number of balanced sign vectors of length $k$: $$C_{k,k/2}=\frac{k !}{(k/2)!(k/2)!}.$$ 
The set of all possible columns $\mathcal{C}(B)^{\mathrm{I}
}$ for each matrix $B \in \mathcal{Z}_{m,z,k}$ in this case can be written as 
\begin{align*}
\mathcal{C}(B)^{\mathrm{I}}
&=\left\{\begin{pmatrix} \mathbf{0}_{m-k} \\ t \end{pmatrix} : t \in \{+1, -1\}^k,\right.\\
&\quad \left. \left. \sum_{i=0}^{k-1} t_i\right. =0 \right\}.
\end{align*}

\item \underline{Case 2:}  
At least one zero appears in the tail of the last column of $B$. The orthogonality constraint from Equation \eqref{refeq2} remains applicable, requiring the sum of the tail entries to be zero.
The non-zero entries in the tail must form a balanced vector of length $s_{tail}$. $z_{tail}$ denotes the number of zeros in the tail. Such a configuration is only possible when $s_{tail}$ is even. The number of such vectors is $$C_{s_{tail},s_{tail}/2}=\frac{s_{tail}!}{(s_{tail}/2)!(s_{tail}/2)!}.$$ In summary, the vectors $t$ generated in this case are all possible length-$k$ vectors whose tail contains exactly $z_{tail}$ zeros and $s_{tail}$ non-zero entries arranged so that the sum of the non-zero elements is zero.
\\ \\
The head, like the tail, is composed of zero and non-zero entries. However, unlike the tail, the non-zero values in the head are not required to form balanced vectors. The only constraints are the prescribed positions of the zeros and the fact that all non-zero entries are equal to $\pm1$. 
If $z_{head}$ denotes the number of zeros in the head and $s_{head}=m-k-z_{head}$ denotes the number of non-zero entries, the composition of the head must satisfy $$z_{head}+z_{tail}=m-k \ and \ s_{head}+s_{tail}=k.$$
The head therefore consists of all $2^{s_{head}}$ possible assignments of $\pm1$ to the non-zero positions, with the remaining $z_{head}$ positions fixed at zero.
\\ \\
To combine the head and the tail, all possible pairs of head and tail vectors are considered. Mathematically, this can be expressed as the Cartesian product of their respective sets: $$\mathcal{C}(B)^{\mathrm{II}
}=\left\{\begin{pmatrix}
    h \\ t
\end{pmatrix}:h \in \mathcal{H},t \in \mathcal{T}\right\}.$$
where the set of valid head vectors $\mathcal{H}$ is defined by $$\mathcal{H}=\{h \in \{-1,0,1\}^{m-k}:\#\{i:h_i=0\}=z_{head}\}.$$ and the set of valid tail vectors $\mathcal{T}$ is 
\begin{align*}
\mathcal{T}&=\big \{t  \in \{-1,0,1\}^k: \\&\quad \sum_{i=0}^{k-1}t_i=0, \#\{i:t_i=0\}=z_{tail}\big \}. \end{align*}
\end{itemize}
The set of candidate columns is thus defined as $$\mathcal{C}(B)=\mathcal{C}(B)^{\mathrm{I}
} \cup \mathcal{C}(B)^{\mathrm{II}
}.$$
For each matrix $B \in \mathcal{Z}_{m,z,k}$, candidate matrices are formed by concatenating the first $z-1$ columns of $B$ with each vector $c \in \mathcal{C}(B)$. If $B_{z-1}$ denotes the sub-matrix of $B$ consisting of its first $z-1$ columns, the candidate matrices are given by $M=[B_{z-1}|c]$ where $c \in \mathcal{C}(B)$. The matrices $M$ are included in the collection $\mathcal{W}_{m,z,k}$ if they satisfy  the orthogonality condition $M'M=kI_z$, where $I_z$ is the $z \times z$ identity matrix and if they are not isomorphic to any matrix already in $\mathcal{W}_{m,z,k}$.
\\ \\
To formalize this, two matrices are considered isomorphic if one can be obtained from the other by permuting rows, permuting columns and flipping the signs of entire rows or columns. Isomorphism is assessed by representing each matrix as a graph, with vertices and edges capturing its combinatorial structure. Graph isomorphism algorithms are then applied to determine whether there exists a vertex relabeling, that does not alter the graph’s structure. Two graphs (as well as the corresponding matrices) are isomorphic if such a relabeling exists.
\\ \\
Let $M\in\{-1,0,1\}^{m \times z}$ be a matrix of size $m \times z$. The graph $G\left(M\right)$ is constructed to encode $M$ in a way that captures row and column permutations, as well as sign flips of rows and columns. A natural first approach is to associate a distinct vertex with each row and each column, connecting them according to the non-zero entries of the matrix. However, this approach does not distinguish between $+1$ and $-1$ entries, so row and column sign flips are not properly represented. In other words, a representation with only one vertex per row and column cannot capture sign changes through vertex relabeling alone. 
To address this limitation, a bipartite graph representation is adopted with positive and negative vertices for each row and column. Details on the construction of the bipartite graph corresponding to a matrix $M$, as well as an illustration of the impact of row and column permutations and sign flips, can be found in Appendix \ref{app:ap2}.
\\ \\
Graph isomorphism is commonly used in experimental design because it efficiently identifies structurally equivalent designs by exploiting inherent symmetries, as highlighted by \cite{Shrivastava2010-ey}. By representing designs as graphs with vertices for experimental units and edges for relationships, automorphisms reveal interchangeable units, so only one representative per equivalence class needs to be evaluated. This avoids redundant computations and significantly reduces computational time compared to purely matrix-based methods.
\section{Results}
This section provides an overview of the enumeration of weighing designs, the computational limits encountered, the strategy to construct larger designs beyond the enumerated cases and the evaluation of OMARS designs derived from the various weighing designs constructed. Section 4.1 reports the number of weighing designs obtained through the enumeration procedure and discusses the computational limits encountered. Section 4.2 describes how larger weighing designs can be generated beyond the enumerated cases using an alternative construction strategy. Section 4.3 introduces the evaluation criteria used to measure the quality of OMARS designs and examines the efficiency of several representative OMARS designs.
\subsection{Enumeration Results} 
\subsubsection{Enumeration of Weighing Designs}
The enumeration was carried out for weighing designs with an even weight $k$ (only even $k$ values are possible as explained in Section 3.2) and a relatively small number of zeros (i.e., a small value of $m-k$). For each value of $m-k$, the number of rows $m$ was progressively increased. The enumeration procedure for each pair $(m,m-k)$ continued until no further columns satisfying the design constraints could be generated or until the computation became infeasible. In cases where the computation became infeasible, the procedure did not complete all iterations even after several days.
\\ \\
A clear structural trend emerged from this process. For a fixed $m$ and $m-k$, the number of non-isomorphic designs initially increases as additional columns are added. This growth continues until the number of columns is about half the number of rows. Beyond this point, however, the number of non-isomorphic designs decreases steadily as the number of columns approaches the number of rows. This pattern is clearly visible in Table \ref{tab:rc-single}, which reports the number of non-isomorphic designs obtained for each combination of number of rows, $m$, and number of columns, $z$. A similar trend was noted in \cite{Conf} for conference designs, which is consistent with the fact that conference designs form a subclass of weighing designs.
\\ \\
This pattern can be explained by the interplay between combinatorial richness and orthogonality constraints. When the number of columns is small relative to the number of rows, there are many ways to extend the design while maintaining the required orthogonality conditions. Consequently, many non-isomorphic designs appear in that case. As the number of columns grows, the orthogonality conditions become increasingly restrictive, reducing the number of admissible extensions. Eventually, when the number of columns is close to the number of rows, the space of valid designs becomes very narrow, leading to fewer (or sometimes no) non-isomorphic designs.
\\
\begin{table*}[!h]
\centering
\caption{Number of non-isomorphic weighing designs as a function of the number of rows $m$ and columns $z$}
\label{tab:rc-single}
\begin{adjustbox}{max width=\textwidth, min width=0.95\textwidth, center}
\small
\setlength{\tabcolsep}{3.2pt}  
\begin{tabular}{@{}c|*{19}{c}@{}}
\toprule
m/z & {2} & {3} & {4} & {5} & {6} & {7} & {8} & {9} & {10} & {11} & {12} & {13} & {14} & {15} & {16} & {17} & {18} & {19} & {20} \\
\midrule
\rowcolor{lightgray} 4  & $2_{\diamond}$ & 1 & 1 & {} & {} & {} & {} & {} & {} & {} & {} & {} & {} & {} & {} & {} & {} & {} & {} \\
5  & $2_{\diamond}$ & $3^{\dagger}$ & {} & {} & {} & {} & {} & {} & {} & {} & {} & {} & {} & {} & {} & {} & {} & {} & {} \\
\rowcolor{lightgray} 6  & $2_{\text{\tiny \#}}$ & $2_{\diamond}$ & 2 & 1 & 1 & {} & {} & {} & {} & {} & {} & {} & {} & {} & {} & {} & {} & {} & {} \\
7  & $2_{\text{\tiny \#}}$ & $4_{\text{\tiny \#}}$ & 2 & 1 & 1 & 1 & {} & {} & {} & {} & {} & {} & {} & {} & {} & {} & {} & {} & {} \\
\rowcolor{lightgray} 8  & $2_{\text{\tiny \#}}$ & $3_{\diamond}$ & 6 & 3 & 3 & 1 & 1 & {} & {} & {} & {} & {} & {} & {} & {} & {} & {} & {} & {} \\
9  & $2_{\text{\tiny \#}}$ & $4_{\diamond}$ & 4 & $2^{\ddagger}$ & {} & {} & {} & {} & {} & {} & {} & {} & {} & {} & {} & {} & {} & {} & {} \\
\rowcolor{lightgray}10 & $2_{\text{\tiny \#}}$ & $3_{\text{\tiny \#}}$ & 8 & 7 & 8 & 4 & 3 & 1 & 1 & {} & {} & {} & {} & {} & {} & {} & {} & {} & {} \\
11 & $2_{\text{\tiny \#}}$ & $5_{\text{\tiny \#}}$ & 7 & 5 & 2 & $1^{\dagger}$ & {} & {} & {} & {} & {} & {} & {} & {} & {} & {} & {} & {} & {} \\
\rowcolor{lightgray}12 & $2_{\text{\tiny \#}}$ & $4_{\diamond}$ & $19_{\diamond}$ & 38 & 81 & 61 & 61 & 27 & 20 & 5 & 5 & {} & {} & {} & {} & {} & {} & {} & {} \\
13 & $2_{\text{\tiny \#}}$ & $5_{\diamond}$ & 9 & 8 & $2^{\dagger}$ & {} & {} & {} & {} & {} & {} & {} & {} & {} & {} & {} & {} & {} & {} \\
\rowcolor{lightgray}14 & $2_{\text{\tiny \#}}$ & $3_{\text{\tiny \#}}$ & 21 & 94 & 216 & 15 & $4^{\dagger}$ & {} & {} & {} & {} & {} & {} & {} & {} & {} & {} & {} & {} \\
15 & $2_{\text{\tiny \#}}$ & $5_{\text{\tiny \#}}$ & 12 & 19 & 11 & $10^{\ddagger}$ & {} & {} & {} & {} & {} & {} & {} & {} & {} & {} & {} & {} & {} \\
\rowcolor{lightgray}16 & $2_{\text{\tiny \#}}$ & $4_{\diamond}$ & 37 & 352 & 3244 & $2210^{\star}$ & $1889^{\star}$ & $1049^{\star}$ & $821^{\star}$ & $472^{\star}$ & $396^{\star}$ & $147^{\star}$ & $67^{\star}$ & $11^{\star}$ & $7^{\star}$ & {} & {} & {} & {} \\
17 & $2_{\text{\tiny \#}}$ & $5_{\diamond}$ & 14 & 26 & 15 & $4^{\ddagger}$ & {} & {} & {} & {} & {} & {} & {} & {} & {} & {} & {} & {} & {} \\
\rowcolor{lightgray}18 & $2_{\text{\tiny \#}}$ & $3_{\text{\tiny \#}}$ & 37 & 884 & $3112^{\star}$ & $3375^{\star}$ & $1157^{\star}$ & $375^{\star}$ & $793^{\star}$ & $597^{\star}$ & $318^{\star}$ & $80^{\star}$ & $18^{\star}$ & {} & {} & {} & {} & {} & {} \\
19 & $2_{\text{\tiny \#}}$ & $5_{\text{\tiny \#}}$ & 17 & 51 & 66 & $71^{\ddagger}$ & {} & {} & {} & {} & {} & {} & {} & {} & {} & {} & {} & {} & {} \\
\rowcolor{lightgray}20 & $2_{\text{\tiny \#}}$ & $4_{\diamond}$ & $57_{\diamond}$ & $603^{\star}$ & $5694^{\star}$ & $2236^{\star}$ & $1312^{\star}$ & $649^{\star}$ & $912^{\star}$ & $1046^{\star}$ & $1321^{\star}$ & $1225^{\star}$ & $1078^{\star}$ & $696^{\star}$ & $746^{\star}$ & $374^{\star}$ & $144^{\star}$ & $22^{\star}$ & $8^{\star}$ \\
21 & $2_{\text{\tiny \#}}$ & $5_{\diamond}$ & 19 & 69 & 135 & $198^{\ddagger}$ & {} & {} & {} & {} & {} & {} & {} & {} & {} & {} & {} & {} & {} \\
\rowcolor{lightgray}22 & $2_{\text{\tiny \#}}$ & $3_{\text{\tiny \#}}$ & 52 & $2301^{\star}$ & $8510^{\star}$ & $3729^{\star}$ & $2567^{\star}$ & $860^{\star}$ & $332^{\star}$ & $164^{\star}$ & $70^{\star}$ & $19^{\star}$ & $5^{\star}$ & {} & {} & {} & {} & {} & {} \\
23 & $2_{\text{\tiny \#}}$ & $2_{\diamond}$ & 4 & $59^{\ddagger}$ & {} & {} & {} & {} & {} & {} & {} & {} & {} & {} & {} & {} & {} & {} & {} \\
\rowcolor{lightgray}24 & $2_{\text{\tiny \#}}$ & $4_{\diamond}$ & 72 & $488^{\star}$ & $13808^{\star}$ & $1708^{\star}$ & $6128^{\star}$ & $651^{\star}$ & $49^{\star}$ & $14^{\star}$ & $3^{\star}$ & {} & {} & {} & {} & {} & {} & {} & {} \\
\bottomrule
\end{tabular}
\end{adjustbox}
\end{table*} \\
The rows highlighted in grey in Table~\ref{tab:rc-single} correspond to even numbers of rows, for which each column contains exactly two zeros. Rows shown in white correspond to odd numbers of rows, where each column contains three zeros. A superscript $\dagger$ or $\ddagger$ indicates that no additional columns can be added while still satisfying the design constraints. The two symbols refer to distinct causes for this limitation:
\begin{itemize}
    \item $\dagger$: no further orthogonal designs exist that meet the specified constraints;
    \item $\ddagger$: there are no more valid zero-pattern extensions allowing a larger design.
\end{itemize}
The latter cause occurs only for an odd number of rows. This suggests that constructing valid zero-pattern extensions is more difficult for an odd number of runs than for an even number, resulting in fewer feasible zero-patterns.
A superscript $\star$ indicates that the corresponding number of designs was obtained through partial rather than complete enumeration.
\\ \\ The results in Table~\ref{tab:rc-single} also allow a direct comparison with previous enumeration studies.
The complete enumeration in \cite{Nunez_Ares2020-ig} was limited to designs with up to 5 columns. For 6 and 7 columns, they used a partial enumeration. As shown in Table \ref{tab:rc-single}, the current enumeration extends considerably further, reaching up to 12 columns for complete enumeration and up to 20 columns with partial enumeration.
\\ \\
Some numbers in the table carry a subscript $\#$ or a subscript $\diamond$. This indicates that, for the corresponding design size, certain weighing designs become strong OMARS designs after foldover augmentation. Naturally, this only happens for small numbers of columns, since it is impossible to estimate all main and interaction effects independently when there are many factors and the number of runs is limited. The difference between the two subscripts is that designs with subscript $\#$ can estimate all effects, including the quadratic effects, whereas those with subscript $\diamond$ are strong OMARS designs that cannot estimate all quadratic effects.
\subsubsection{Partial Enumeration}
\begin{figure}[!h]
    \centering
    \fbox{%
    \includegraphics[width=0.95\linewidth]{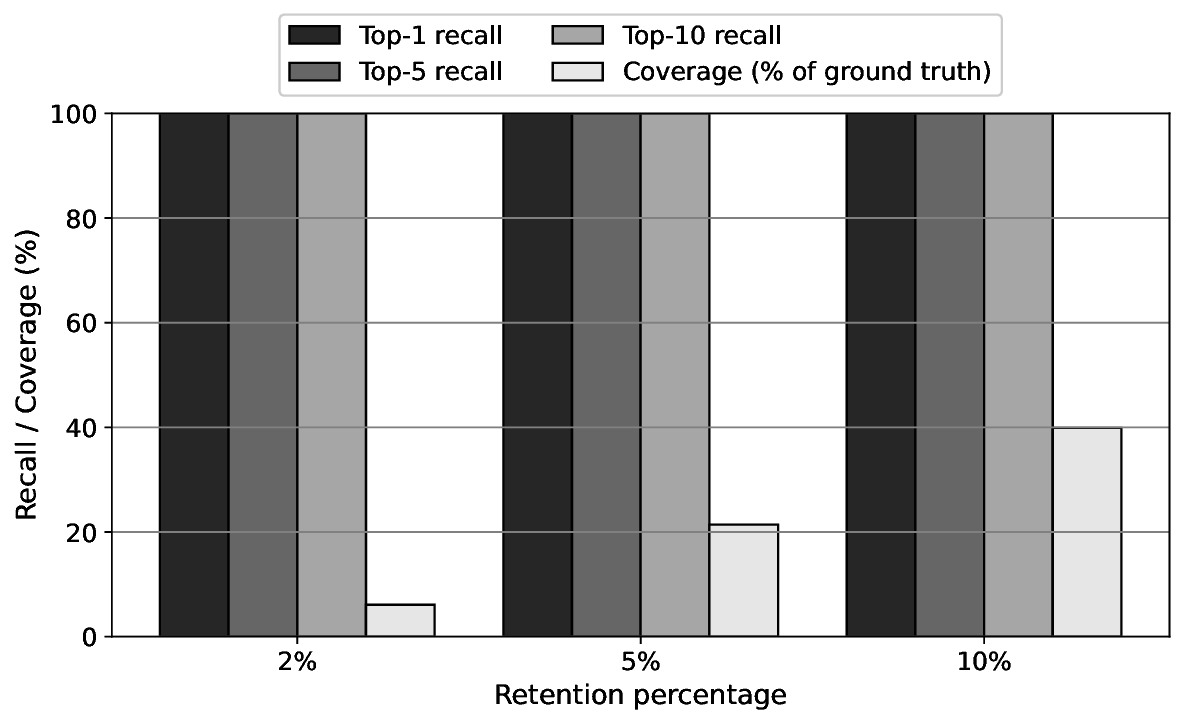}}
    \caption{Top-$N$ recall and coverage of the partial enumeration
    procedure at $m=16$}
    \label{fig:partial-validation-m16}
\end{figure}

Complete enumeration becomes increasingly demanding as the number of rows and columns grows.
Consequently, an alternative strategy was required to extend the catalog beyond the range that can be completely enumerated. To address this, a partial enumeration approach was used to generate only the
most promising designs. At each step, the designs in $\mathcal{W}_{m,z-1,k}$
were ranked from best to worst and only the top fraction of designs was retained for extension to $z$ columns. The ranking criterion utilized here is the one used to compare all OMARS designs later in this paper and is defined in Section 4.3.1. The retention percentage varied with $(m, z)$ and is reported in
Appendix \ref{app:6}.
\\ \\
This procedure is justified by a structural property of the designs. When a
design is extended by an additional column, the correlations among the effects already present remain unchanged. The extension only adds new correlations involving the effects of the new column. A $(z-1)$-column design ranked below another therefore starts the next step with a disadvantage that the extension itself cannot undo. The share of inherited correlations grows with $z$, making it structurally harder for a low-ranked design to overtake a top-ranked one when extended with an extra column. The partial enumeration procedure is therefore expected to retain the designs that lead to the best final designs. A formal proof of this property and a quantitative analysis are given in Appendix \ref{app:7}.
\\ \\
To validate this expectation empirically, the partial enumeration procedure was tested at $m=16$. At this size, complete enumeration produced 3244
non-isomorphic weighing designs at $z=6$ and partial enumeration
takes over from $z=7$ onward (see Table~\ref{tab:rc-single}). This
makes the transition from $z=5$ to $z=6$ the most informative
within-reach test of whether the partial enumeration filter preserves
the designs that dominate the final ranking.
\\ \\
The validation was performed as follows. The complete enumeration at $z=5$
contains 352 non-isomorphic weighing designs. The partial-enumeration filter was applied to this set: the designs were ranked from best to worst using the criterion of Section 4.3.1 and the top fraction was retained. The retained designs were then extended to $z=6$ through the column-by-column construction and isomorphism filtering described in Section 3. The
resulting partial-enumeration set was then compared to the
3244 designs obtained by the complete enumeration at $z=6$. Three retention percentages were tested: 2\%, 5\% and
10\%.
\\ \\
Two metrics are reported. The first is the top-$N$ recall, i.e., the fraction of the $N$ best designs from the complete enumeration that are present in the partial set, up to isomorphism. The second is the coverage by design count, i.e., the number of designs in
the partial set expressed as a fraction of the 3244 designs from the complete
enumeration.
\\ \\
Figure~\ref{fig:partial-validation-m16} reports the comparison. The single
best OMARS design, the top-5 designs and the top-10 designs are all recovered at all three retention percentages. The coverage by design count grows with the retention percentage, from 6.1\% at 2\% to 39.9\% at 10\%. Even the tightest retention percentage of 2\% therefore preserves the entire top-10 set.
\\ \\
Complete enumeration returns all non-isomorphic weighing designs and thus guarantees that the optimal OMARS designs are found. Partial enumeration can in principle discard a design that would have led to a top-tier final design after further extension, but the structural argument and the validation at $m = 16$ indicate that this loss is most likely limited in practice. A similar principle underlies the partial enumeration of \cite{schoen2017} for larger two-level orthogonal screening designs.
\subsubsection{Computational Characteristics}
Beyond the enumeration results themselves, it is also informative to examine the computational effort required to obtain them. The enumeration algorithm was implemented in C. This choice was motivated by the large combinatorial search spaces arising in the enumeration procedure. In particular, the low-level implementation in C enables tight control over memory allocation and compact storage of intermediate candidates, which is essential when exploring large search trees. This also allows fast orthogonality checks and efficient pruning, resulting in practical runtimes for complete enumeration up to 24 tests.
\\ \\
Table \ref{tab:rc-single2} reports the incremental runtime required to find all non-isomorphic weighing designs with one extra column, assuming that the previous case has already been computed and stored. The runtime increases very sharply with the number of rows $m$, with computations moving from seconds to minutes, hours and even days for larger $m$. The effect of the number of columns $z$ is not monotonic. The runtime is largely proportional to the number of designs that must be extended at each step
(Table~\ref{tab:rc-single}), which implies that the runtime is maximal when the number of columns is
about half the number of rows. For $m = 16$, for example, the extension step
to $z = 6$ required 9.6 days. These results illustrate both the feasibility and the computational limitations of complete enumeration, motivating the partial enumeration described in Section 4.1.2 and the concatenation strategy proposed in Section 4.2 for larger OMARS constructions.
\\ \\
All computations were performed on the same machine equipped with an Intel® Core™ i9-10980XE CPU @ 3.00 GHz and 125 GB RAM, under identical conditions. This ensures that the reported runtimes are directly comparable across configurations and that the observed increase in computational time can be attributed solely to the growth in problem complexity.
\begin{figure}[!h]
    \centering
    \fbox{\includegraphics[width=\linewidth]{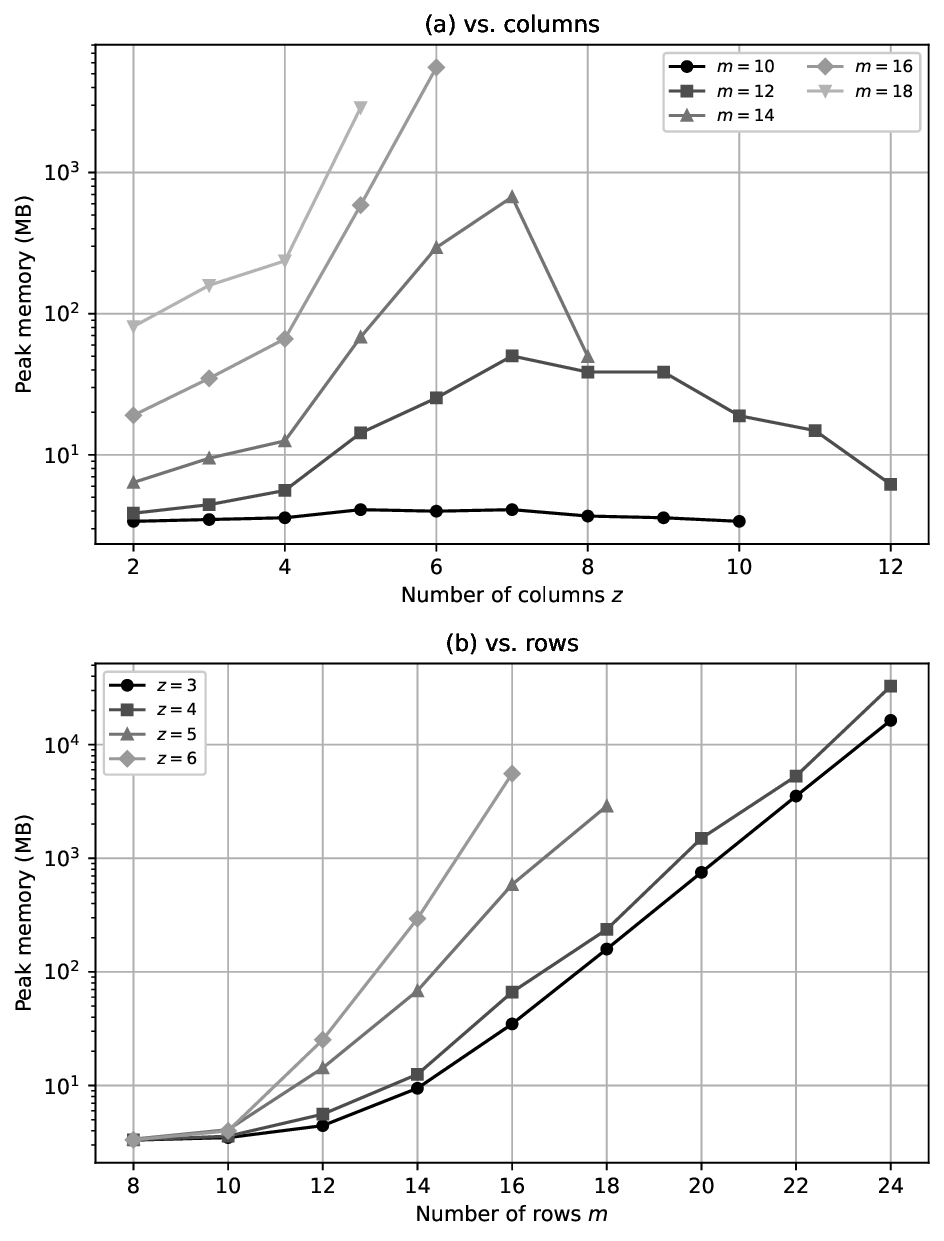}}
    \caption{Peak memory of a single column-extension step versus the
number of columns $z$ (a) and the number of rows $m$ (b)}
    \label{fig:memory}
\end{figure}
\\ \\
Runtime is only one aspect of computational scalability. The peak memory required by the enumeration was also profiled. The peak memory of a single column-extension step is the maximum amount of working memory occupied while the candidate designs (the matrices
$M = [B_{z-1} \mid c]$ of Section 3) are stored before isomorphism filtering.
Figure~\ref{fig:memory} shows how this peak memory varies with $m$ and $z$, and two patterns emerge.
\begin{table*}[!t]
\centering
\caption{Incremental runtime for fully enumerated weighing designs with $m$ rows and $z$ columns}
\label{tab:rc-single2}
\scriptsize
\setlength{\tabcolsep}{4pt}
\begin{tabular}{@{}c|*{11}{c}@{}}
\toprule
$m$/$z$ & 2 & 3 & 4 & 5 & 6 & 7 & 8 & 9 & 10 & 11 & 12 \\
\midrule
\rowcolor{lightgray}
4  & $< 1$ sec & $< 1$ sec & $< 1$ sec & {} & {} & {} & {} & {} & {} & {} & {} \\
5  & $< 1$ sec & $< 1$ sec & {} & {} & {} & {} & {} & {} & {} & {} & {} \\
\rowcolor{lightgray}
6  & $< 1$ sec & $< 1$ sec & $< 1$ sec & $< 1$ sec & $< 1$ sec & {} & {} & {} & {} & {} & {} \\
7  & $< 1$ sec & $< 1$ sec & $< 1$ sec & $< 1$ sec & $< 1$ sec & $< 1$ sec & {} & {} & {} & {} & {} \\
\rowcolor{lightgray}
8  & $< 1$ sec & $< 1$ sec & $< 1$ sec & $< 1$ sec & $< 1$ sec & $< 1$ sec & $< 1$ sec & {} & {} & {} & {} \\
9  & $< 1$ sec & $< 1$ sec & $< 1$ sec & $< 1$ sec & {} & {} & {} & {} & {} & {} & {} \\
\rowcolor{lightgray}
10 & $< 1$ sec & $< 1$ sec & $< 1$ sec & $< 1$ sec & $< 1$ sec & $< 1$ sec & $< 1$ sec & $< 1$ sec & $< 1$ sec & {} & {} \\
11 & $< 1$ sec & $< 1$ sec & $< 1$ sec & $< 1$ sec & $< 1$ sec & $< 1$ sec & {} & {} & {} & {} & {} \\
\rowcolor{lightgray}
12 & $< 1$ sec & $< 1$ sec & $5$ sec & $32$ sec & $1.9$ min & $3.5$ min & $2.6$ min & $57$ sec & $11$ sec & $1$ sec & $<1$ sec \\
13 & $4$ sec & $2$ sec & $<1$ sec & $<1$ sec & $<1$ sec & {} & {} & {} & {} & {} & {} \\
\rowcolor{lightgray}
14 & $5$ sec & $4$ sec & $28$ sec & $8.5$ min & $32$ min & $1.7$ min & $4$ sec & {} & {} & {} & {} \\
15 & $25$ sec & $15$ sec & $2$ sec & $3$ sec & $<1$ sec & $<1$ sec & {} & {} & {} & {} & {} \\
\rowcolor{lightgray}
16 & $32$ sec & $35$ sec & $5.5$ min & $4.8$ hr & $9.6$ day & {} & {} & {} & {} & {} & {} \\
17 & $2.5$ min & $55$ sec & $10$ sec & $14$ sec & $5$ sec & $3$ sec & {} & {} & {} & {} & {} \\
\rowcolor{lightgray}
18 & $4.1$ min & $1.3$ min & $3.7$ min & $1.7$ hr & {} & {} & {} & {} & {} & {} & {} \\
19 & $15.2$ min & $7.1$ min & $45$ sec & $2.7$ min & $1.2$ min & $1.6$ min & {} & {} & {} & {} & {} \\
\rowcolor{lightgray}
20 & $28.7$ min & $19.4$ min & $4.3$ hr & {} & {} & {} & {} & {} & {} & {} & {} \\
21 & $1.4$ hr & $25.2$ min & $3.9$ min & $14.5$ min & $7.7$ min& $14.9$ min & {} & {} & {} & {} & {} \\
\rowcolor{lightgray}
22 & $2.9$ hr & $38.9$ min & $2.5$ hr & {} & {} & {} & {} & {} & {} & {} & {} \\
23 & $7.5$ hr & $3$ hr & $16.6$ min & $46.4$ min & {} & {} & {} & {} & {} & {} & {} \\
\rowcolor{lightgray}
24 & $16.6$ hr & $7.8$ hr & $11.6$ hr & {} & {} & {} & {} & {} & {} & {} & {} \\
\bottomrule
\end{tabular}
\end{table*}
For a fixed number of rows (panel~a), the memory does not grow monotonically
with the number of columns. It increases with $z$ toward a maximum near
$z \approx m/2$ and then declines. This mirrors the evolution of the number
of non-isomorphic designs as a function of $z$ in
Table~\ref{tab:rc-single}. For smaller row counts, where the complete
enumeration was performed for all possible values of $z$, this pattern is
perfectly visible in panel~a. For larger row counts, where the complete
enumeration was only feasible for small values of $z$, only the increasing
part of the evolution in peak memory usage is visible.
\\ \\
For a fixed number of columns (panel~b), the peak memory required grows
steeply with the number of rows. The curves for the different column
counts are close to parallel on the logarithmic scale. This indicates a similar growth rate with $m$ for each value of $z$, with each additional column shifting the curve upward. 
\\ \\
The peak memory thus reflects the combinatorial
size of the enumeration. It is governed by the number of designs
being processed and the number of candidate extensions generated for
each. Both increase with the size of the design, i.e., $m$ and $z$. Across the fully enumerated configurations, the peak memory reaches
at most about 32\,GB. Complete
enumeration is therefore constrained by runtime rather than by
memory.
\\ \\
The computational characteristics of the proposed approach can also be viewed in relation to the OMARS design enumeration of \cite{Nunez_Ares2020-ig}, involving an integer program solved with a general-purpose solver. The two approaches rely on different algorithms and hardware and are not directly comparable in terms of runtime or memory. They nonetheless operate within the same class of computational resources, a single workstation with gigabytes of memory and runtimes ranging from seconds to a few days. Within that class, the present method reaches the substantially larger designs reported above.
\subsection{Concatenating Weighing Designs}
Concatenation is a method for constructing larger weighing designs beyond what enumeration alone can produce. As demonstrated in Appendix \ref{app:ap3}, two weighing designs with the same number of factors can be concatenated to produce a new weighing design by stacking them vertically. The resulting design includes all runs from both original designs and its weight is the sum of the weights of the two original designs.
\\ \\ 
This property can be used to construct larger designs. Moreover, if two weighing designs can be concatenated into a larger weighing design, this new, larger design can in turn be concatenated with yet another design, and so on, potentially without limit. This recursive property allows for generating progressively larger designs. Note, however, that while this approach allows the number of rows to increase freely, the number of columns remains fixed. 
\\ \\
It is interesting to determine which sizes of designs can be obtained from the enumeration in Table \ref{tab:rc-single} by exploiting this recursive property. 
Let $\mathcal{M}=\{m_1, m_2, \ldots, m_s \}$ be the set of initial row counts corresponding to designs with column count $z$. Then the set of attainable total rows counts, denoted by $\mathcal{S}_z$, is the set of all non-negative integer combinations of the elements of $\mathcal{M}$: $$\mathcal{S}_z=\{n_1m_1+n_2m_2+\ldots+n_sm_s |n_i \in\mathbb{N}_0\},$$ where each $n_i$ is a non-negative integer representing the number of times a design with $m_i$ rows is used in the concatenation. If the greatest common divisor $g$ of the row sizes in $\mathcal{M}$ is larger than 1, the only attainable totals in $\mathcal{S}_z$ are multiples of $g$. If the greatest common divisor is 1, there is no divisibility restriction and all sufficiently large integers can be represented as sums of the elements of $\mathcal{M}$. In practice, for sets with many numbers, the attainable totals rapidly become dense, meaning that essentially all integers above a certain threshold can be achieved \citep{Ramirez_Alfonsin2005-py}.
\\ \\
By concatenating the designs in Table \ref{tab:rc-single}, the best weighing designs with 25 to 78 runs were generated. 
The concatenated designs were chosen in such a way that the number of factors that could be obtained was maximized, while minimizing the number of designs required, emphasizing larger designs whenever possible. For each feasible run size, the best design was generated for each attainable number of factors.
\subsection{OMARS designs}
This section examines the performance of the OMARS designs generated. Section 4.3.1 introduces the evaluation criteria used to identify the best designs within the OMARS framework. Section 4.3.2 presents several representative large OMARS designs and reports their efficiency.
\subsubsection{Evaluation Criteria}
OMARS designs are intended to investigate main effects, two-factor interactions and quadratic effects. In these designs, the correlation matrix of the columns from the model matrix (other than the intercept column) naturally separates into distinct blocks, each corresponding to a different type of combination of model term, reflecting the overall structure of the design.
For a design with $r$ factors, the correlation matrix has dimension $$(r+\frac{r(r-1)}{2}+r) \times (r+\frac{r(r-1)}{2}+r),$$
corresponding respectively to the $r$ main effects, the $r(r-1)/2$ two-factor interactions and the $r$ quadratic effects. 
This matrix can be partitioned into sub-matrices, each representing correlations within or between groups of effects:
\begin{itemize}
    \item The main–main block is an identity matrix $I_r$, indicating complete orthogonality among main effects.
    \item The main–interaction and main–quadratic blocks are zero matrices, since OMARS designs enforce orthogonality between main effects and second-order terms.
    \item The interaction–interaction (II) and quadratic–quadratic (QQ) blocks are square matrices that contain the correlations (aliasing) among the two-factor interactions and among the quadratic effects, respectively. In contrast, the interaction–quadratic (IQ) block is a rectangular matrix that represents the correlations between the interaction and quadratic terms. None of these blocks are generally diagonal or zero, as they capture the residual aliasing among second-order effects.
\end{itemize}
The structure of the correlation matrix can be represented as
$$\begin{bmatrix}
I_{r \times r} & 0_{r \times \tfrac{r(r-1)}{2}} & 0_{r \times r} \\[6pt]
0_{\tfrac{r(r-1)}{2} \times r} & II_{\tfrac{r(r-1)}{2} \times \tfrac{r(r-1)}{2}} & IQ_{\tfrac{r(r-1)}{2} \times r} \\[6pt]
0_{r \times r} & IQ^{\top}_{r \times \tfrac{r(r-1)}{2}} & QQ_{r \times r}
\end{bmatrix}.$$
In this matrix, II represents the correlations among the two-factor interactions, 
QQ represents the correlations among the quadratic effects and 
IQ represents the correlations between the interactions and the quadratic terms. The structure of the matrix shows the orthogonality of the main effects and their separation from the second-order terms, while indicating the locations of the residual correlations among the second-order terms.
\\ \\
Minimizing aliasing among effects is crucial for estimating each effect as independently and as precisely as possible and it enhances the interpretability of the effects. Achieving this requires minimizing the entries of the blocks II, IQ and QQ, thereby reducing the correlations among two-factor interactions, between interactions and quadratic effects and among quadratic effects. The choice of the order of prioritization depends on the intended use of the design. In this paper, the following order has been adopted: the II block is minimized first, followed by the IQ block and finally the QQ block. Interactions are the most numerous of the second-order effects, so controlling the block II has the largest impact on precision and interpretability. Correlations between interactions and quadratics (IQ) are minimized next, reflecting their intermediate importance. Quadratic effects are fewer, making QQ correlations the least critical. This prioritization produces a minimally aliased structure among the
second-order effects.
\\ \\ 
Two quality measures have been selected to get a summary of all correlations of a certain type: the Average Absolute Correlation ($AAC$) and the Root-Mean-Square ($RMS$) correlation. The $AAC$ provides a measure of the typical strength of correlations within a specific block of the correlation matrix. Let $B$ denote a block of interest (II, IQ or QQ) with $n_1^{(B)}$ rows and $n_2^{(B)}$ columns. The $AAC$ for block $B$ is defined as $$AAC^{(B)}=\frac{1}{n_1^{(B)}n_2^{(B)}}\sum_{i=1}^{n_1^{(B)}}\sum_{j=1}^{n_2^{(B)}}|r_{ij}^{(B)}|,$$ where $r_{ij}^{(B)}$ is the correlation between the effect corresponding to the $i$-th row and the effect corresponding to the $j$-th column of block $B$.
\\ \\
The $RMS$ correlation emphasizes larger correlations that may have a disproportionate impact on the estimation precision. It is defined as $$RMS^{(B)}=\sqrt{\frac{1}{n_1^{(B)}n_2^{(B)}}\sum_{i=1}^{n_1^{(B)}}\sum_{j=1}^{n_2^{(B)}}\left( r_{ij}^{(B)} \right)^2 }$$
By squaring correlations before averaging, the $RMS$ gives more weight to strong correlations, which can inflate variances and reduce interpretability even if most correlations are small. These two measures are complementary: the $AAC$ captures the typical level of correlations, while the $RMS$ values highlight large correlations that could dominate variance inflation. To compare two OMARS designs, a geometric mean ($GM$) of the $AAC$ and $RMS$ values is computed for each block. For a given block $B$, the $GM$ is defined as $$GM^{(B)}=\sqrt{AAC^{(B)} \cdot RMS^{(B)}}.$$
The $GM$ combines the overall magnitude of the correlations ($AAC$) with the impact of large correlations ($RMS$) into a single summary measure. Unlike a simple sum, the geometric mean balances the two contributions multiplicatively, so a design with a very small $AAC$ but an exceptionally large $RMS$, or vice versa, will not appear artificially favorable. This makes $GM$ a more sensitive and fair measure of aliasing within a block.
\\ \\
All OMARS designs constructed from the weighing designs in Sections 4.1 and 4.2 were first compared based on their $GM$ value for the II block, since minimizing correlations among interactions was prioritized. If the $GM$s for the II block were equal, the comparison proceeded to the IQ block and finally to the QQ block, reflecting the decreasing priority of these blocks in controlling aliasing.
\subsubsection{Large OMARS designs}
\begin{figure}[!h]
\centering
\fbox{
\begin{tabular}{@{}c@{}c@{}c@{}c@{}}
\subf{\includegraphics[width=31 mm]{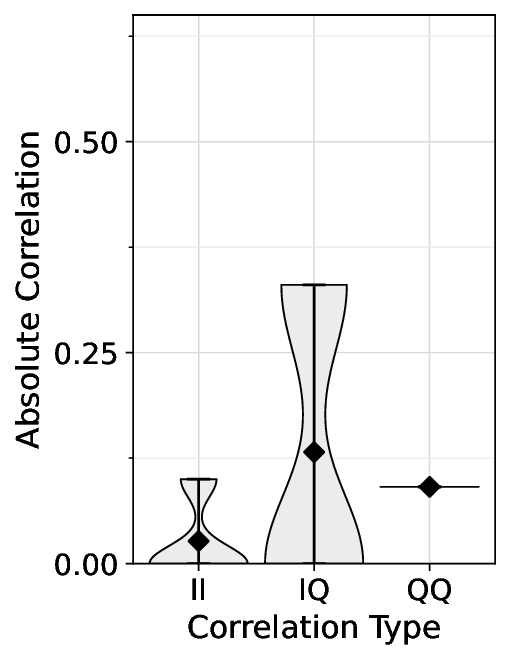}}
     {(a) 96 x 5} &
\subf{\includegraphics[width=31 mm]{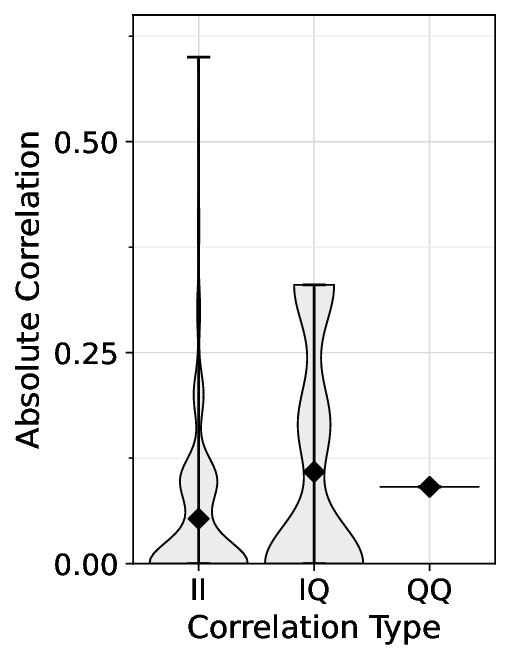}}
     {(b) 96 x 9} \\
\subf{\includegraphics[width=31 mm]{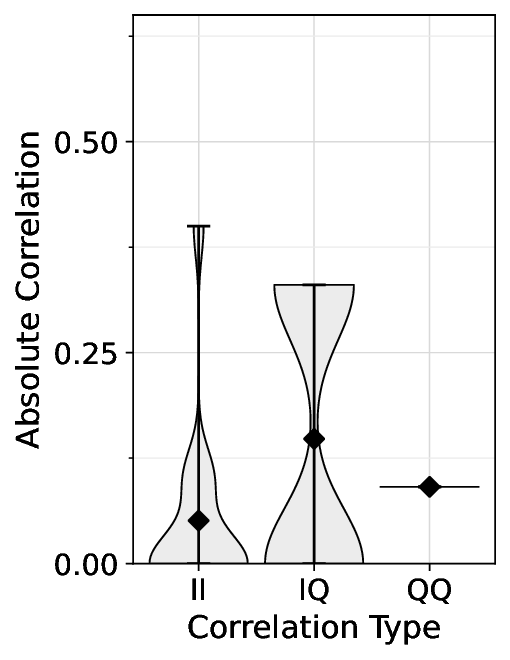}}
     {(c) 144 x 7} &
\subf{\includegraphics[width=31 mm]{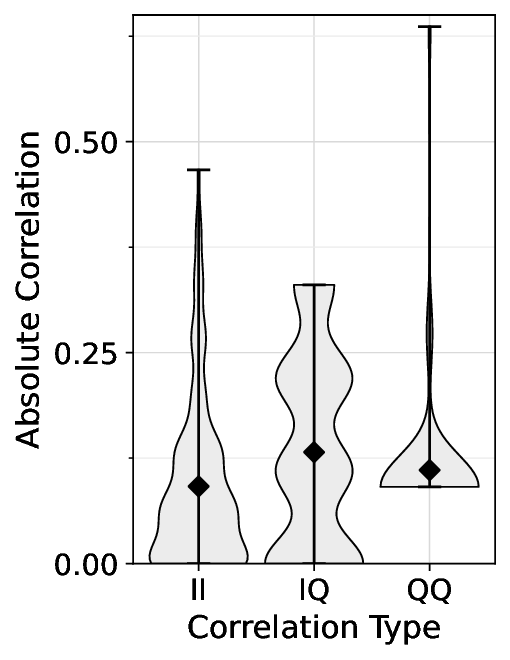}}
     {(d) 144 x 11}
\end{tabular}}
    \caption{Violin plots of the absolute second-order correlations (II, IQ and QQ) for two OMARS designs with 96 runs (panels a and b) and two with 144 runs (panels c and d). Correlations that are zero by construction are omitted.}
    \label{fig:fourfigs}
\end{figure}

\begin{figure}[!h]
\centering
\fbox{
\begin{tabular}{@{}c@{}c@{}c@{}c@{}}
\subf{\includegraphics[width=31 mm]{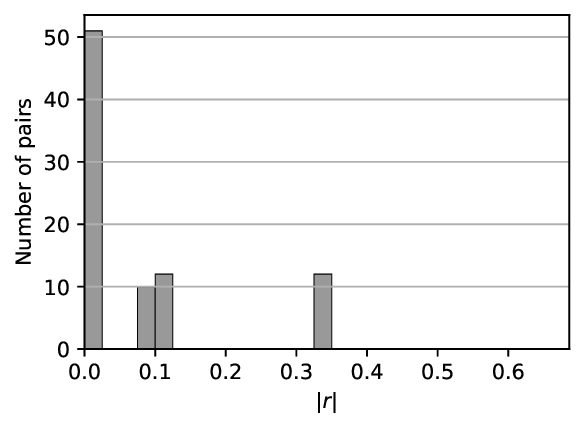}}
     {(a) 96 x 5} &
\subf{\includegraphics[width=31 mm]{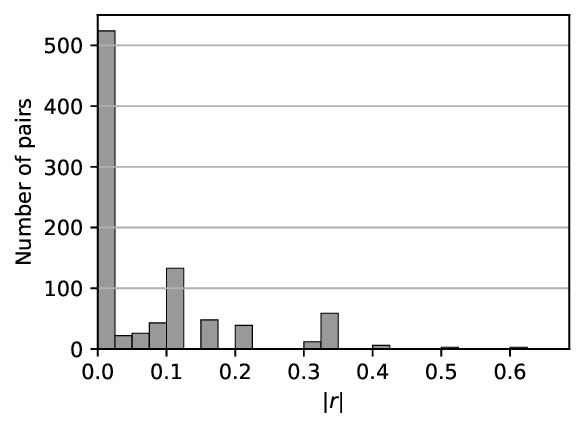}}
     {(b) 96 x 9} \\
\subf{\includegraphics[width=31 mm]{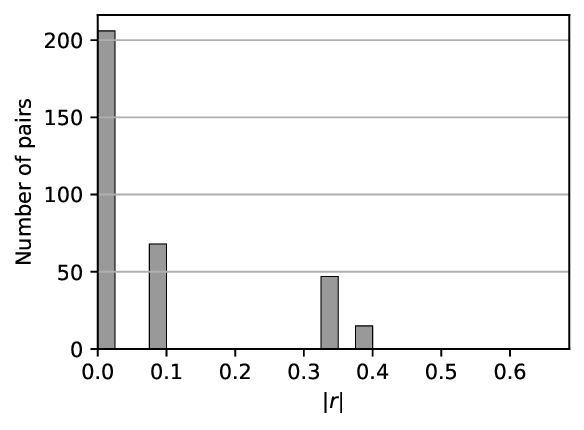}}
     {(c) 144 x 7} &
\subf{\includegraphics[width=31 mm]{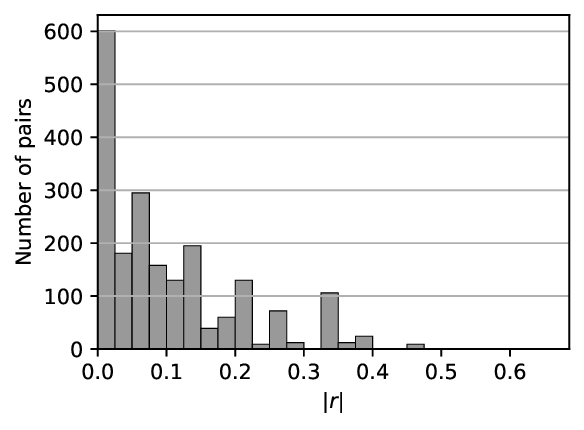}}
     {(d) 144 x 11}
\end{tabular}}
    \caption{Histograms of all second-order correlations for the four OMARS designs of Figure~\ref{fig:fourfigs}. Correlations that are zero by construction are omitted.}
    \label{fig:hist}
\end{figure}

\begin{figure}[!h]
    \centering
    \fbox{
    \includegraphics[width=1\linewidth]{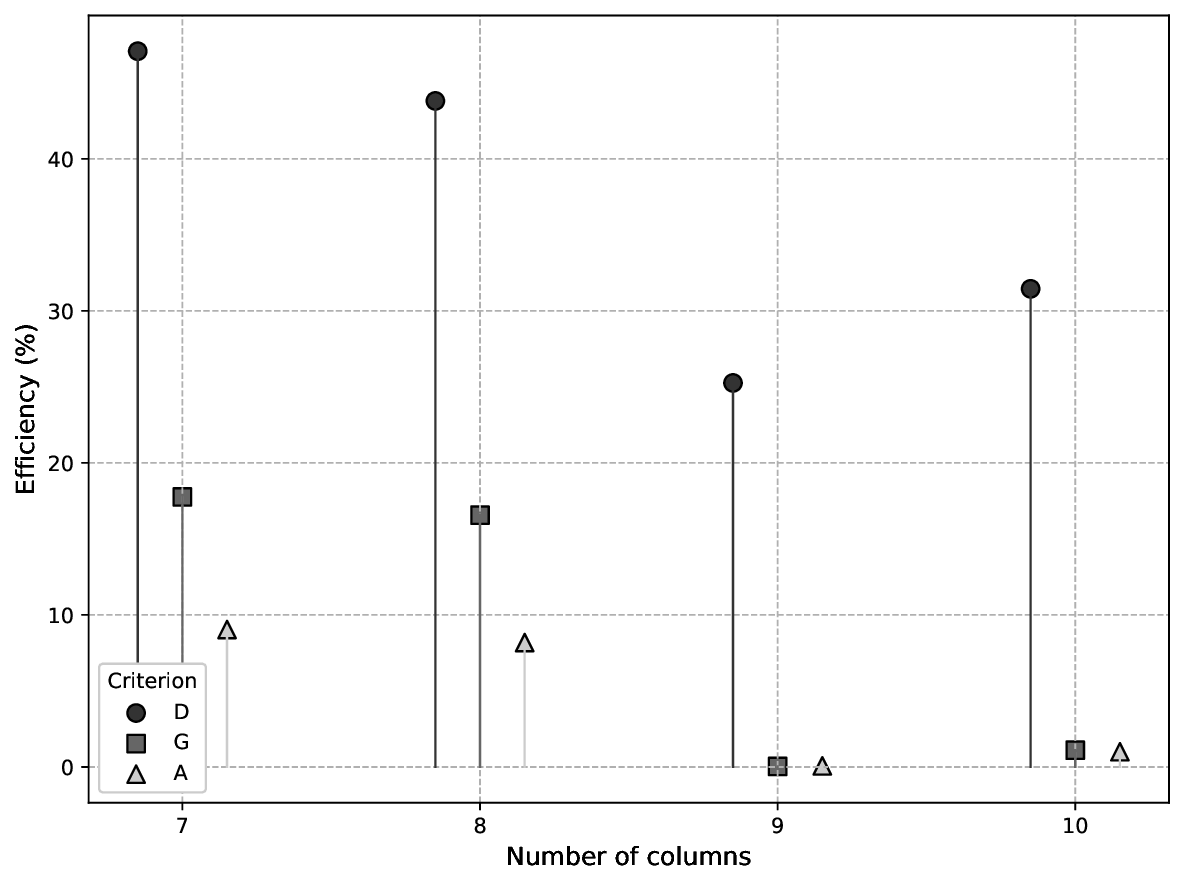}}
    \caption{Design Efficiency for four 128-run OMARS designs}
    \label{figD}
\end{figure}

\begin{figure}[!h]
    \centering
    \fbox{
    \includegraphics[width=1\linewidth]{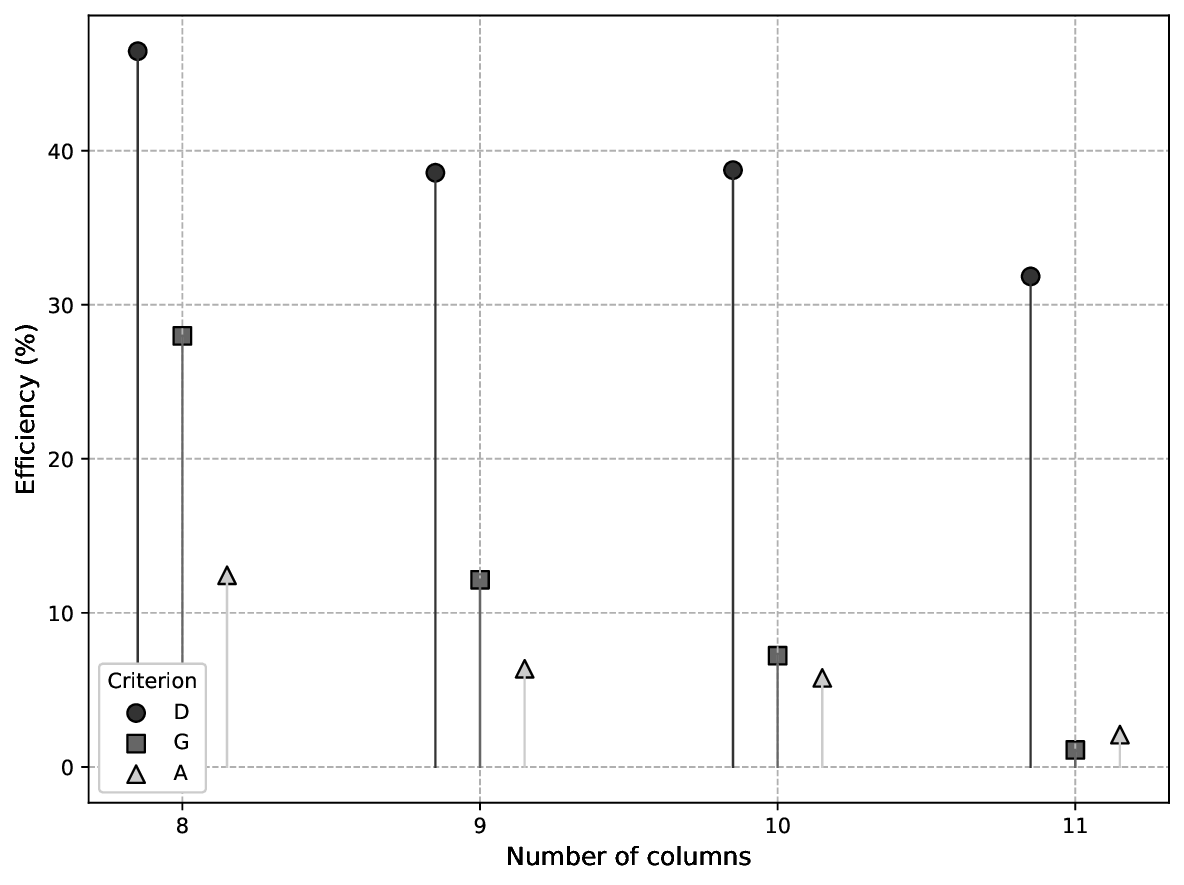}}
    \caption{Design Efficiency for four 156-run OMARS designs}
    \label{figG}
\end{figure}

Based on the weighing designs generated in Sections 4.1 and 4.2, the best OMARS designs for each run size were created. Each OMARS design was obtained by folding over the corresponding weighing designs to produce designs with 24 to 156 runs, with the number of factors depending on the combination of weighing designs used in the concatenation. Some of the resulting designs are discussed below.
\begin{figure*}[!h]
\setlength{\abovecaptionskip}{1pt}
\centering
\fbox{%
\begin{minipage}{0.90\linewidth}
\centering
    \includegraphics[scale=0.55]{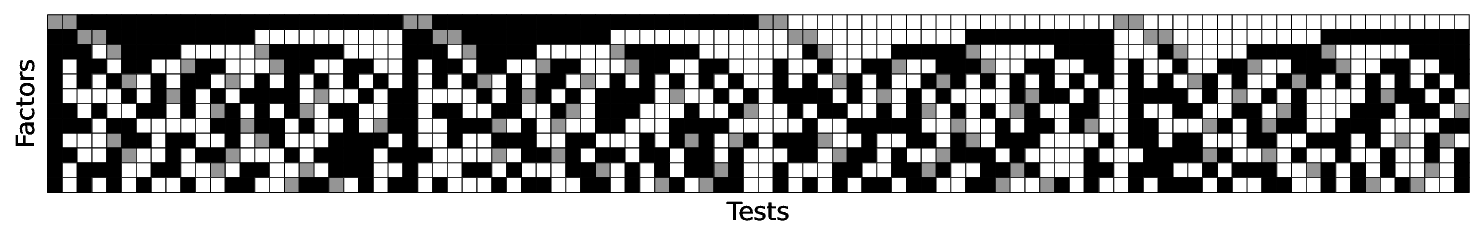}
    \caption{Heatmap of a 96 x 12 OMARS design}
    \label{fig:design96}
\end{minipage}
}
\end{figure*}
\\ \\
Figure \ref{fig:design96} shows an example of an OMARS design with 96 rows and 12 columns, where each column contains eight zeros. The 96-run size is particularly convenient in practice, as it matches the 96-well plate format that is standard in pharmaceutical high-throughput screening \citep{Macarron2011}. The full 96-run experiment can therefore be carried out on a single plate without splitting runs across multiple plates. This design was constructed by concatenating two of the three available 24 × 12 weighing designs and then applying a foldover to obtain an OMARS design. To obtain that particular design, the three unique concatenation pairs were evaluated and the best design according to the $GM$ criterion described in Section 4.3.1 was selected. The design is displayed as a heat-map, with black representing the factor level 1, grey representing the factor level 0 and white representing the factor level $-$1. For visualization, the matrix is transposed so that factors appear along the horizontal axis and runs along the vertical axis. The foldover structure is immediately apparent: splitting the design vertically down the middle reveals that the right half is a color-inverted version of the left half.
\\ \\
As explained in Section 4.3.1, the quality of the large OMARS designs can be
first assessed using the absolute values of the three nonzero correlation
types (II, IQ and QQ). Figure~\ref{fig:fourfigs} presents violin plots of
these correlations for four large OMARS designs, which are the best ones
found for their respective sizes. Each violin plot illustrates the
distribution of absolute correlations, highlighting the minimum and maximum
values, as well as the mean, which is indicated by a diamond. The correlations between each interaction and the quadratic effects of its own two factors are excluded, as they are zero by construction in every OMARS design.
\\ \\
The designs with 96 rows were generated using the procedure described earlier for generating and selecting the design in Figure \ref{fig:design96}. The 144-run designs were constructed by combining three 24-run weighing designs. The best 72-run combination of these three designs was then selected and folded over to obtain an OMARS design with 144 rows.
\\ \\
The II correlations consistently show the lowest mean, indicating that aliasing between interactions is generally the smallest. However, for three of the four designs under study, the II correlations exhibit substantial variability, due to a limited number of high correlations between the interactions. In three of the four designs, the QQ correlations are all identical. For the 144 × 11 design, however, they take three distinct values, the largest of which (0.64) exceeds the maximum II correlation. Finally, there are many high IQ correlations overall, reflecting that correlations between interactions and quadratic effects are a major source of aliasing in these designs.
\\ \\
Increasing the number of columns generally worsens the aliasing by raising
the overall mean for all three correlation types (II, IQ, QQ), as well as the variability and the frequency of high correlation values for the II
correlations. For the IQ correlations, the violin plots become wider in the
middle, indicating that there are more intermediate correlations for larger
numbers of factors than for smaller numbers of factors. For the QQ
correlations, the increase in columns has no effect for the 96-run example,
while it has a noticeable effect for the 144-run example. The reason is that the QQ correlation between two quadratic effects depends
only on the number of rows in which both factors are set to zero
simultaneously. In the 96-run designs and in the 144-run design with 7
columns, every row of the design contains at most one zero. No two factors
are therefore ever at their middle level in the same row, and all QQ
correlations coincide. In the 144-run design with 11 columns, some rows
contain two zeros. The pairs of factors that meet at their middle level in
such rows have different QQ correlations, so three distinct QQ values appear.
\\ \\
A closer inspection of the high correlations reveals that they
are concentrated in a small number of factor pairs rather than
spread across many. Figure~\ref{fig:hist} shows the histograms of
all second-order correlations, comprising the three nonzero
correlation types (II, IQ and QQ), for the four large OMARS designs of Figure~\ref{fig:fourfigs} (96 × 5, 96 × 9, 144 × 7 and 144 × 11). These
designs comprise 85, 918, 336 and 2035 such correlations,
respectively. In all four cases, the majority of the correlations
lie close to zero, with a small number of outliers in the tail.
\\ \\
For the 144 × 11 OMARS design, the maximum absolute II correlation is 0.47 (visible in Figure~\ref{fig:hist}(d)), while the average across all 1485 pairwise II correlations is only 0.09, a ratio of about five. The top five II correlations all reach this maximum value and involve specific pairs of interactions such as $(X_1 X_{10},\ X_9 X_{11})$ and $(X_3 X_4, \ X_5 X_{11})$. The 96 × 9 OMARS design shows the same
pattern more sharply, with a maximum II correlation of 0.60 against
an average of 0.05, and the top correlations involving interactions
among the factors $X_1, \ X_5,\ X_7$ and $X_9$ exclusively.
\\ \\
The IQ and QQ blocks behave differently from the II block. The maximum IQ correlation is approximately 0.33 across the four
designs considered (visible as a small spike in each panel of Figure~\ref{fig:hist}). The QQ correlations take only a small number of distinct values within each design, and in three of the four
designs of Figure~\ref{fig:fourfigs}, they are even identical across all
factor pairs. This consistency in the high IQ and QQ
correlations is due to the zero patterns of the underlying
weighing designs, which are common to all designs with given
$m$ and $z$ values.
\\
\\
The infrequent occurrence of high II correlations has a direct practical
implication for the use of the catalog of OMARS designs constructed: a user
with prior knowledge of which two-factor interactions are critical for their
application can assign the factors involved in these interactions
to columns that do not appear in the list of the most strongly correlated
interaction pairs. For the 144 × 11 OMARS design, for example, columns
indexed by $\{X_2, X_6, X_8\}$ do not appear in any of the most strongly
correlated interaction pairs, making them safe positions for critical factor
assignment. For the 96 × 9 OMARS design, columns $\{X_3, X_6\}$ play the same role. Since the IQ and QQ outliers are common to all factor columns, they cannot be mitigated by factor-to-column reassignment. The II aliasing of critical factors, by contrast, can be substantially reduced by an appropriate choice of assignment.
\\ \\
To understand how the three types of correlations affect statistical quality, the D-, G- and A-efficiencies were evaluated for four 128-run designs and for four 156-run OMARS designs for second-order models including main effects, two-factor interactions and quadratic effects. D-efficiency reflects the overall precision of the coefficient estimates by measuring the determinant of the information matrix, A-efficiency measures the average variance of the estimated coefficients and G-efficiency captures the worst-case prediction variance. 
In all three cases, higher values indicate a more statistically efficient design. 
\\ \\
The 128-run designs were created by concatenating two 20-run weighing designs with a 24-run weighing design. The best combination was selected and then folded over to produce the final 128-run OMARS designs. Similarly, the 156-run designs were generated by concatenating three 20-run weighing designs with one 18-run design. The best combination was retained and subsequently folded over to yield the final 156-run OMARS designs. Figures \ref{figD} and \ref{figG} visualize the designs’ D-, G- and A-efficiencies, using lollipop plots to show how each criterion varies with the number of factors.
\\ \\
The plots clearly reveal the inherent trade-off between model complexity and statistical efficiency. For a fixed number of runs, all three efficiencies (D-, A- and G-) decline as the number of factors increases, since more columns imply more parameters to estimate with the same amount of information and, inevitably, more aliasing among the second-order effects (see, for instance, Figures~\ref{fig:fourfigs} and \ref{fig:hist}). For a given number of factors, designs with a larger number of runs (e.g., 156) consistently achieve higher efficiencies than designs with fewer runs (e.g., 128), demonstrating their capacity to support more complex models.
\\ \\
Quantitatively, the D-efficiency starts near 50 \% and decreases to around 30 \% as the number of factors increases. The G-efficiency begins around 20 \% and approaches zero, while the A-efficiency starts near 10 \% and also tends toward zero for large numbers of factors. These trends are consistent with the results reported by \cite{Nunez_Ares2020-ig} for small OMARS designs. The near-zero G- and A-efficiencies for larger column counts show that predictive and average estimation precision for the full second-order model decline rapidly as model complexity increases. Since there is generally effect sparsity, the low G- and A-efficiencies for the full model are not worrying: the efficiencies for submodels are much higher.
\\ \\
The best-known three-level alternatives to OMARS designs are DSDs. For this reason, 96-run OMARS designs involving 5, 8, 9 and 12 factors were also contrasted with DSDs of the same size. For $k = 5$ and $k = 8$, where the full second-order model is estimable under
both designs, the OMARS designs achieve higher D-, A- and G-efficiencies. At
$k = 9$ and $k = 12$, neither design permits estimation of the full second-order model, as both are foldover designs. In terms of aliasing, the OMARS designs also achieve lower average II correlations
at $k = 5$, $8$ and $9$ and lower average IQ correlations at all four factor
counts, whereas the DSDs achieve a lower average QQ correlation. This lower QQ
correlation does not, however, indicate a more precise estimation of the
quadratic effects: each DSD factor takes its middle level in only two of the 96 runs, so the
quadratic effects are estimated far less precisely than under the OMARS designs. Under the GM criterion used throughout this paper, the
OMARS designs rank ahead of the DSDs at $k = 5$, $8$ and $9$. At $k = 12$ the two designs are nearly tied, with the DSD marginally ahead. The detailed comparison between the $96$-run OMARS designs and DSDs is reported in Appendix~\ref{app:dsd}.
\\ \\
The D- and A-efficiencies in Figures~\ref{figD} and \ref{figG} demonstrate
that large OMARS designs can estimate full second-order models involving a
substantial number of factors, achieving efficiencies comparable to those of
smaller OMARS designs. However, as discussed by \cite{Stallrich2025-vc},
foldover designs improve certain design properties, but they reduce the
number of independent pieces of information available for estimation. As a
result, the total number of factors that can be included while still allowing estimation of all second-order effects is limited. This limitation is reflected in the maximum number of columns $c_{max}$ that can be retained in an $r$-run foldover design while being able to estimate the full second-order response surface model,
$${c_{max}=\left\lfloor \frac{-1+\sqrt{4r-7}}{2}\right\rfloor}.$$
That number only grows very gradually with the run size $r$. The detailed
derivation of the $c_{max}$ formula is provided in
Appendix \ref{app:ap5}. All designs in Figures \ref{fig:fourfigs}, \ref{figD} and \ref{figG} have fewer than $c_{max}$ columns.
\\ \\
The dependence of $c_{max}$ on the run size is summarized in Table \ref{tab:cmax}. A 96-run foldover design supports at most 9 factors for a full second-order model. A 156-run foldover design supports at most 11 factors. Users facing this constraint can either reduce the number of
factors below $c_{max}$ to retain full estimability, or accept that not all
second-order effects can be estimated simultaneously.
\\
\begin{table}[!h]
\centering
\caption{Values of $c_{max}$ as a function of $r$}
\label{tab:cmax}
\begin{tabular*}{\linewidth}{@{\extracolsep{\fill}}cc@{}}
\toprule
Run size $r$ & $c_{max}$ \\
\midrule
$48 \le r \le 57$   & 6 \\
$58 \le r \le 73$   & 7 \\
$74 \le r \le 91$   & 8 \\
$92 \le r \le 111$  & 9 \\
$112 \le r \le 133$ & 10 \\
$134 \le r \le 157$ & 11 \\
\bottomrule
\end{tabular*}
\end{table}
\vspace{-1 cm}
\\
Note that $c_{max}$ is only relevant to experiments in which the full second-order response surface model is to be estimated.
\cite{Goos2025-gu} explained that OMARS designs are excellent choices for
performing screening and response surface optimization in a single
experiment. In that case, the number of factors is generally substantially larger
than $c_{max}$ and advanced model selection techniques are then used to
identify the active effects. Generic sparse estimation methods such as the Lasso \citep{Lasso1995} and the Dantzig selector \citep{PHOA2009} can be used for this purpose, but methods tailored to screening experiments are more attractive because they respect the effect heredity principle and offer alternative interpretations of the data. Examples are the mixed integer optimization approach of \citet{Vazquez2021-jq} and the method of \citet{Hameed2023-fh}.
\section{Conclusion}
In this article, an enumeration procedure for weighing designs has been developed, extending the classical concept of weighing matrices to designs with fewer columns. The procedure enables the complete enumeration of designs with up to 24 rows in a limited amount of time, covering different numbers of factors and weights corresponding to two or three zeros per column. This article also provides a partial enumeration procedure for designs with a larger number of factors for a given number of rows.
A concatenation method was also proposed to construct larger weighing designs than those directly obtainable through enumeration. For each number of
factors, the concatenation method supports a wide range of numbers of rows.
\\ \\
Finally, all weighing designs obtained were combined with their foldovers to construct large OMARS designs. For these large OMARS designs, the correlations among model effects were evaluated together with the D-, G- and A-efficiencies. The results also demonstrate that large OMARS designs can estimate second-order models involving a substantial number of factors with efficiencies comparable to those of smaller OMARS designs. 
\\ \\
Three limitations of the proposed methodology were identified throughout the paper. First, for any given run size, the number of factors for which the full second-order model can be estimated is bounded above by $c_{max}$ because the foldover structure consumes degrees of freedom that would otherwise be available for estimating the second-order model. Fortunately, due to the existence of advance model selection techniques, estimability of the full second-order model is no longer required to identify active main effects, interaction effects and quadratic effects. Second, complete enumeration is computationally feasible only for designs with up to 24 runs. Beyond this threshold, the design space expands too rapidly for exhaustive enumeration. Third, partial enumeration extends the catalog beyond the reach of complete enumeration but does not guarantee that the very best design for every $(m, z)$ pair is included. However, the empirical validation in Section 4.1 supports the conclusion that the quality loss incurred by the partial enumeration is likely to be negligible. 
\\ \\
These limitations also suggest several directions for future research. First, extending the catalog to larger run sizes most likely requires a different strategy. Instead of enumerating all admissible designs, the focus would have to move to identifying high-quality near-OMARS designs for specified run sizes. Metaheuristic optimisation methods are particularly well suited to this setting. For instance, simulated annealing and evolutionary algorithms have an established track record in the construction of high-quality experimental designs. More recent learning-based and reinforcement-learning search strategies may offer additional opportunities. These approaches sacrifice the completeness guarantee in exchange for scalability to substantially larger design spaces. A second direction is to extend the enumeration algorithm
to non-uniform-precision designs,
mixed-level designs combining quantitative and categorical factors
\citep{mixedOMARS}, and blocked OMARS designs \citep{blockedOMARS}. These subclasses introduce additional structural
constraints, which would require new isomorphism filters and modified
column-extension rules. A third direction is to increase the number of columns of the constructed designs through other concatenation techniques, such as the combination of two weighing matrices described by \cite{Goos2025-gu}. The catalog of non-isomorphic weighing designs produced in this paper provides the building blocks for such constructions. Finally, a fourth direction for future research is to investigate optimization strategies for constructing concatenated OMARS designs. Existing optimization frameworks developed for other classes of experimental designs (e.g., \citet{Vazquez2026}) could provide a promising starting point for extending such approaches to OMARS designs.
\\ \\
A complete catalog of the best weighing designs obtained from the enumeration, the partial enumeration and the concatenation method, as well as the best OMARS designs generated from them, is available from the authors and the corresponding C code is also available upon request.
\newpage
\appendix
\refstepcounter{section}
\section*{Appendix A}
\label{app:apA}
Before generating OMARS designs, it is useful to prove why combining a weighing design with its foldover yields an OMARS design. Consider a weighing design $W \in \{-1,0,1\}^{m\times z}$ with weight $k$ and define
$$D=\begin{bmatrix} \phantom{-}W \\ -W \end{bmatrix} \in \{-1,0,1\}^{2m \times z},$$ where
$$d_i=\begin{bmatrix} \phantom{-}w_i \\ -w_i \end{bmatrix},
$$
and $d_i$ and $w_i$ denote column $i$ of $D$ and $W$, respectively. 
\\ \\
An OMARS design is a three-level design in which:
\begin{itemize}
    \item all main effects are mutually orthogonal,
    \item main effects are orthogonal to all two-factor interactions,
    \item main effects are orthogonal to all quadratic effects.
\end{itemize}
Each requirement is verified below.
\begin{enumerate}
    \item Since each column of $D$ is the concatenation of $w_i$ and its negation, all entries of $D$ remain in $\{-1,0,1\}$. Thus $D$ is a three-level design.
    \item Distinct main effects in $D$ are orthogonal. Consider two distinct columns $d_i$ and $d_j$. Their inner product is
    \begin{align*}
d_i' d_j
&=
\begin{bmatrix}
w_i' & (-w_i)'
\end{bmatrix}
\begin{bmatrix}
w_j \\ -w_j
\end{bmatrix} \\
&=
w_i' w_j + (-w_i)'(-w_j) \\
&=
2 (w_i' w_j).
\end{align*}
    Because \(W\) is a weighing design, every pair of distinct columns $w_i$ and $w_j$ is orthogonal:
    \[
    w_i'w_j = 0.
    \]
    Therefore,
    \[
    d_i'd_j = 0,
    \]
    showing that all main effects in \(D\) are mutually orthogonal. Note that the orthogonality of the main effects also follows from the results in Appendix \ref{app:ap3}, since an OMARS design is a concatenation of the weighing designs $W_1=W$ and $W_2=-W$.
    \item Each main effect is orthogonal to every two-factor interaction. Let \(\odot\) denote the element-wise product. A two-factor interaction column for
    factors \(i\) and \(j\) is represented by \(d_i \odot d_j\). The foldover structure of $D$ implies that
    \[
    d_i \odot d_j
    =
    \begin{bmatrix}
    w_i \odot w_j \\
    (-w_i)\odot(-w_j)
    \end{bmatrix}
    =
    \begin{bmatrix}
    w_i \odot w_j \\
    w_i \odot w_j
    \end{bmatrix}.
    \]
    
    The inner product between the main effect column \(d_h\) and the interaction column 
    \(d_i \odot d_j\) is
    \begin{align*}
    d_h'(d_i \odot d_j)
    &=
    \begin{bmatrix}
    w_h' & (-w_h)'
    \end{bmatrix}
    \begin{bmatrix}
    w_i \odot w_j \\
    w_i \odot w_j
    \end{bmatrix} \\
    &=
    w_h'(w_i \odot w_j) - w_h'(w_i \odot w_j)
    = 0.
    \end{align*}
    Thus every main effect is orthogonal to every two-factor interaction.
    \item Each main effect is orthogonal to every quadratic effect. A quadratic column for factor \(i\) is represented by
\[
d_i^{(2)} = d_i \odot d_i
=
\begin{bmatrix}
w_i \odot w_i \\
(-w_i)\odot(-w_i)
\end{bmatrix}
=
\begin{bmatrix}
w_i^{2} \\
w_i^{2}
\end{bmatrix}.
\]

The inner product between a main effect column \(d_h\) and this quadratic column is
\begin{align*}
d_h' d_i^{(2)}
&=
\begin{bmatrix}
w_h' & (-w_h)'
\end{bmatrix}
\begin{bmatrix}
w_i^{2} \\
w_i^{2}
\end{bmatrix} \\
&=
w_h' w_i^{2} - w_h' w_i^{2}
= 0.
\end{align*}

Hence, each main effect is orthogonal to every quadratic effect.
\end{enumerate}
As a conclusion, $D$ is an OMARS design, since the required orthogonality conditions hold, due to the foldover construction and the structure of the weighing design.
\newpage
\refstepcounter{section}
\section*{Appendix B}
\label{app:apB}
The goal is to show that the column-by-column enumeration procedure is correct and complete. This proof has two parts: the first part shows that every valid $m \times z$ design arises as an extension of a unique $(z-1)$-column representative and the second part shows that all constraints on candidate columns are preserved under the transformations in $G$.
\\ \\
For every matrix $A \in \mathcal{W}_{m,z-1,k}$,
the algorithm generates exactly one representative from each isomorphism class of
\(m \times z\) weighing designs under the transformations in
\begin{align*}
G
&= \{ \text{row permutations, column permutations,} \\
&\quad \text{row sign flips, column sign flips} \}.
\end{align*}

First, assume that the enumeration
of \((z-1)\)-column designs is complete up to isomorphism. Let \(D\) be an arbitrary \(m \times z\) weighing design. Remove an arbitrary column
\(c\) from \(D\) and denote the remaining matrix by \(D^{\ast}\). There exists a
representative
\[
A \in \mathcal{W}_{m,z-1,k}
\]
and an isomorphism transformation \(g \in G\) such that
\[
g(A) = D^{\ast}.
\]

Apply the inverse transformation \(g^{-1}\) to the removed column \(c\):
\[
c^g = g^{-1}(c).
\]

If the transformation preserves the constraints imposed on candidate columns
(verified below), then \(c^g\) satisfies exactly the same constraints as the columns
generated during the extension of \(A\). Hence, \(c^g\) is a valid candidate column that the
procedure will consider. Appending it to $A$ produces
\[
[A \mid c^g].
\]
Applying \(g\) reproduces the original design:
\[
g([A \mid c^g]) = D.
\]

Since \(D\) was arbitrary, every valid \(m \times z\) design appears (up to
isomorphism) as an extension of some representative
\[
A \in \mathcal{W}_{m,z-1,k}.
\]
Because the enumeration procedure discards only columns that are equivalent under transformations in $G$, each isomorphism class contributes exactly one
representative. This establishes completeness.

It remains to show that all constraints on the candidate column are preserved under the transformations in \(G\). Let \(c\) be any column and let
\(c^g = g^{-1}(c)\). Consider each operation:

\begin{itemize}
    \item \textbf{Row permutations} simply reorder entries and therefore preserve the
    number of zeros, the counts of \(+1\) and \(-1\) and all inner products with the
    existing columns.
    
    \item \textbf{Column permutations} reorder only the existing columns of \(A\),
    preserving all existing zero patterns and preserving all
    orthogonality relations with the new column.

    \item \textbf{Row sign flips} multiply the same row of every column by \(-1\),
    preserving zero positions and preserving all inner products (both vectors involved in an inner product undergo
    the same row flip, so the product is unchanged).

    \item \textbf{Column sign flips} multiply all entries of a column by \(-1\),
    preserving orthogonality, feasibility of sign patterns and all zero patterns.
\end{itemize}

Since every operation in \(G\) preserves every constraint used in the
extension step, the transformed column \(c^g = g^{-1}(c)\) satisfies exactly the same
conditions as the columns generated for the representative \(A\).
\newpage
\refstepcounter{section}
\section*{Appendix C}
\label{app:ap2}
Each row and column of a weighing design $M$ is represented by two vertices in the graph required for checking the isomorphism (in Stage 2 of the enumeration algorithm in Section 3.2) : one for its positive version and one for its negative version. This produces a graph that is bipartite with respect to row and column vertices, while each part is subdivided into positive and negative nodes. This representation allows encoding of all three matrix entry values $\{-1, 0, 1\}$ using edges corresponding to non-zero entries. Formally, the vertex set of the graph $G\left(M\right)$ is defined as: 
\begin{displaymath}
  V=\{r_i^+,r_i^-\left|0\le i<m\}\ \cup\ \{c_j^+,c_j^-\right|0\le j<z\} \nonumber
\end{displaymath}
\noindent
where $r_i^+$ and $ r_i^-$ represent the positive and negative versions of row $i$, and $c_j^+$ and $c_j^-$ represent the positive and negative versions of column $j$. \\ \\
The edge set $E$ encodes both the non-zero entries of the matrix and the correspondence between the positive and negative vertex of each row and column, ensuring that sign flips are properly represented.
Specifically, the edge set is constructed as follows:
$$
\begin{aligned}
E = & \ \{ (r_i^+, c_j^+), (r_i^-, c_j^-) \mid M_{ij} = 1 \} \\
     &  \ \cup \ \{ (r_i^+, c_j^-), (r_i^-, c_j^+) \mid M_{ij} = -1 \} \\
    & \ \cup \ \{ (r_i^+, r_i^-) \mid 0 \le i < m \} \\
     &  \ \cup \ \{ (c_j^+, c_j^-) \mid 0 \le j < z \}.
\end{aligned}
$$
In this encoding, for every matrix entry $M_{ij}=1$, edges connect vertices of the same sign: $(r_i^+, c_j^+)$ and $ (r_i^-, c_j^-)$. For every matrix entry $M_{ij}=-1$, edges connect vertices of opposite signs: $(r_i^+, c_j^-)$ and $ (r_i^-, c_j^+)$. Flipping the sign of a row or column simply swaps its positive and negative vertices, preserving the overall connectivity pattern, while row or column permutations correspond to relabeling the associated vertices. The edges $(r_i^+, r_i^-)$ and $(c_j^+, c_j^-)$ connect the positive and negative vertices of each row or column, ensuring that they are treated as representing the same row or column under a possible sign flip. \\ \\
Finally, to ensure that only valid transformations are captured during isomorphism testing, the graph is vertex-colored: all row vertices share one color and all column vertices share a second color. These colors are structural constraints: any graph isomorphism is required to preserve vertex colors. Consequently, rows can only be permuted with rows and columns only with columns, while sign flips of both rows and columns are allowed, all without violating the bipartite structure of the graph. The coloring therefore restricts the isomorphism group to transformations that correspond exactly to admissible row and column operations on the matrix.
\\
\begin{figure}[!h]
\setlength{\abovecaptionskip}{1pt}
\fbox{
\centering
    \includegraphics[scale=1.1]{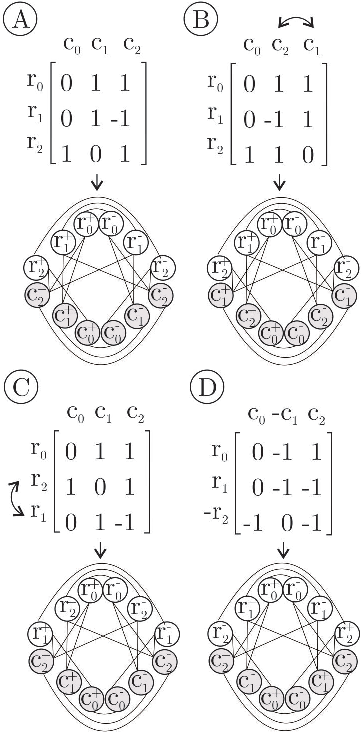}}
    \caption{Example of vertex relabeling process}
    \label{fig:enter-label}

\end{figure}
\\
Figure \ref{fig:enter-label} provides examples of each type of the vertex relabeling corresponding to the allowed matrix transformations. Panel A shows an initial example matrix $M$ of size $3 \times 3$ and its associated graph $G(M)$. Since each row and column is represented by two vertices (positive and negative versions), the graph contains $2 \times 3+2\times 3=12$ vertices, with white representing row vertices and gray representing column vertices.
\\ \\
In Panel B, columns $c_1$ and $c_2$ are permuted, which in the graph corresponds to relabeling the positive vertex of $c_1$ to $c_2^+$ and the positive vertex of $c_2$ to $c_1^+$, as well as relabeling the negative vertex of $c_1$ to $c_2^-$ and the negative vertex of $c_2$ to $c_1^-$. In Panel C, rows $r_1$ and $r_2$ are swapped, which translates in the graph to relabeling the positive vertex of $r_1$ to $r_2^+$ and the positive vertex of $r_2$ to $r_1^+$, along with relabeling the negative vertex of $r_1$ to $r_2^-$ and the negative vertex of $r_2$ to $r_1^-$.
\\ \\
Finally, in Panel D, the second column $c_1$ undergoes a sign flip, which corresponds in the graph to exchanging its positive and negative vertices. The panel also involves a sign flip in the last row, as a result of which the positive and negative vertices of that row are exchanged too. 
\\ \\
In the Panels B, C and D, the connectivity pattern of the graph is preserved, demonstrating that these matrix transformations correspond to color-preserving vertex relabeling in $G(M)$. This justifies why graph isomorphism algorithms can be used to check the isomorphism of the weighing designs.
\newpage
\refstepcounter{section}
\section*{Appendix D}
\label{app:6}
This appendix reports the retention percentages applied at each step
of the partial enumeration procedure described in Section 4.1. At
each step from $z-1$ to $z$ columns, the procedure ranks the designs
in $\mathcal{W}_{m, z-1, k}$ by the GM criterion of Section 4.3.1 and retains the top fraction listed in
Table~\ref{tab:retention} for extension to $z$ columns.
\\ \\
The entries of Table~\ref{tab:retention} are interpreted as follows.
An entry marked ``C'' indicates that complete enumeration was
performed at the corresponding step, so that the partial filter was
not applied and every design from the previous step was carried
forward. A numeric entry gives the percentage of designs retained by
the partial enumeration procedure, where the value 100 indicates that
the partial procedure was applied in an earlier phase but all surviving designs with $z-1$ columns were extended to $z$ columns. A dash indicates a step to which no retention percentage applies. This occurs
either when $z > m$, since a weighing design cannot have more columns than
rows, or when the designs retained at the previous step admitted no valid
one-column extension, so that the enumeration produced no design with $z$
columns.
\\ \\
The transition from complete to partial enumeration occurs earlier
for larger $m$. For $m=16$, complete enumeration is feasible through $z=6$ and for $m=18$ through $z=5$. For $m \geq 20$, the partial procedure must
be invoked from $z=5$ onward, since the design population at that
step is already too large for an exhaustive sweep.
\begin{table}[!h]
\centering
\caption{Retention percentages (\%) used in the partial enumeration}
\label{tab:retention}
\small
\begin{tabular*}{\linewidth}{@{\extracolsep{\fill}}cccccc@{}}
\toprule
$z$ & $m=16$ & $m=18$ & $m=20$ & $m=22$ & $m=24$ \\
\midrule
5   & C      & C      & 5.00   & 10.00  & 1.00   \\
6   & C      & 2.00   & 1.00   & 0.20   & 0.30   \\
7   & 5.00   & 2.00   & 0.20   & 0.10   & 0.01   \\
8   & 10.00  & 2.00   & 1.00   & 0.50   & 1.00   \\
9   & 10.00  & 5.00   & 2.00   & 1.00   & 0.25   \\
10  & 20.00  & 100.00 & 10.00  & 5.00   & 1.90   \\
11  & 30.00  & 100.00 & 15.00  & 30.00  & 100.00 \\
12  & 100.00 & 100.00 & 20.00  & 100.00 & 100.00 \\
13  & 100.00 & 100.00 & 20.00  & 100.00 & --     \\
14  & 100.00 & 100.00 & 25.00  & 100.00 & --     \\
15  & 100.00 & --     & 30.00  & --     & --     \\
16  & 100.00 & --     & 100.00 & --     & --     \\
17  & --     & --     & 100.00 & --     & --     \\
18  & --     & --     & 100.00 & --     & --     \\
19  & --     & --     & 100.00 & --     & --     \\
20  & --     & --     & 100.00 & --     & --     \\
21  & --     & --     & --     & --     & --     \\
22  & --     & --     & --     & --     & --     \\
23  & --     & --     & --     & --     & --     \\
24  & --     & --     & --     & --     & --     \\
\bottomrule
\end{tabular*}
\end{table}
\vspace{- 1cm}
\\
The retention percentage is generally lowest at intermediate values of $z$,
where the number of designs in $\mathcal{W}_{m, z-1, k}$ is largest
and the total cost of the extension step is therefore highest. As $z$ grows further, the number of designs in $\mathcal{W}_{m,z-1,k}$ starts to shrink and a larger fraction can again be retained. In the validation experiment of Section 4.1, retention values of 2\%, 5\% and 10\% were tested
explicitly for the extension of 5-factor designs to 6-factor designs for $m=16$. These values are within the working range used elsewhere in the catalog.
\newpage
\refstepcounter{section}
\section*{Appendix E}
\label{app:7}
This appendix proves that the GM ranking of OMARS designs at
step $z$ structurally influences the ranking at step $z+1$ in favour of
the top-ranked designs retained by the partial enumeration
procedure. The notation follows Section 3 for $\mathcal{W}_{m,z,k}$ and Section 4.3.1 for
the aliasing blocks II, IQ and QQ and the quantities $AAC^{(B)}$, $RMS^{(B)}$
and $GM^{(B)}$. For a weighing design $W \in \mathcal{W}_{m,z,k}$, let $D$ be the OMARS design obtained by folding over $W$.
\\ \\
Let $W \in \mathcal{W}_{m,z,k}$ and let $W' = [W \mid c]$ be a one-column
extension, with OMARS designs $D$ and $D'$. Denote the $j$-th column of $D$ by
$\widetilde{c}_j$. The first $z$ columns of $D'$ are identical to those of $D$,
hence every product $\widetilde{c}_i \odot \widetilde{c}_j$ for
$1 \le i < j \le z$ and every squared column $\widetilde{c}_i \odot
\widetilde{c}_i$ for $1 \le i \le z$ is unchanged. As a consequence, every entry of
II$(D)$, IQ$(D)$ and QQ$(D)$ appears unchanged in the corresponding top-left
sub-matrix of II$(D')$, IQ$(D')$ and QQ$(D')$.
\\ \\
When a design is extended from $z$ to $z+1$ columns, the entries of
the II block split into two groups: those already present at step
$z$ (the inherited entries) and those that involve the newly added
column (the new entries). Let $N^{II}_z = \binom{z}{2}^2$ denote the
number of entries in the II block for the $z$-column design (consistent with
the definition of $AAC^{(II)}$ in Section 4.3.1
), and let
$\rho = N^{II}_z / N^{II}_{z+1}$ be the fraction of entries
inherited at the next step. Because the inherited entries are
unchanged, 
$AAC^{(II)}$ can be rewritten as
\[
AAC^{(II)}(D')
\;=\;
\rho \, AAC^{(II)}(D)
\;+\;
(1 - \rho) \, \overline{|r|}^{\,\mathrm{new}},
\]
where $\overline{|r|}^{\,\mathrm{new}}$ is the average absolute correlation over
the new entries only. The same kind of identity holds for $(RMS^{(II)})^2$,
with $\overline{r^2}^{\,\mathrm{new}}$ in place of
$\overline{|r|}^{\,\mathrm{new}}$. Analogous identities hold for the IQ and QQ blocks. Each $AAC^{(B)}$ and
$(RMS^{(B)})^2$ value at step $z+1$ is therefore a weighted average of its value at
step $z$ and a contribution from the new entries.
\\ \\
Consider two weighing designs $A$ and $B$ at step $z$ whose foldovers differ in $AAC^{(II)}$ by an amount $\Delta > 0$, with $A$ being the better of the two. After both designs have been extended by one column, the gap between their new
$AAC^{(II)}$ values is
\[
\rho \, \Delta
\;+\;
(1 - \rho) \, \bigl(
\overline{|r|}^{\,\mathrm{new}, B'} -
\overline{|r|}^{\,\mathrm{new}, A'} \bigr).
\]
The first part of this expression is the original advantage of $A$ carried over with
weight $\rho$. The second part is the difference between the average absolute correlation over the
new entries of $B$ and over the new entries of $A$, weighted by $1 - \rho$. Since the average absolute correlation is at most one, the largest possible difference caused by the new entries is $1 - \rho$. This bound is attained in the extreme case, where the new correlations of $B$ are all zero and those of $A$ all equal
one. 
For $B$ to overtake $A$ at step $z+1$, the gap between their new $AAC^{(II)}$
values must become negative. In this extreme case, that gap reduces to
$\rho\,\Delta - (1 - \rho)$, which is negative only when
$\rho\,\Delta < 1 - \rho$, or equivalently when
\[
\Delta \;<\; \Delta^* \;=\; \frac{1 - \rho}{\rho}.
\]
Two designs separated by more than $\Delta^*$ in $AAC^{(II)}$ value at
step $z$ therefore cannot swap order at step $z+1$, whatever column
is chosen to extend either of them. Some example values of $\rho$ and
$\Delta^*$ for the II block at several transitions are:
\begin{itemize}
    \item $z=5 \to 6$: \;\; $\rho = 100/225 \approx 0.44$, \;\; $\Delta^* \approx 1.25$;
    \item $z=10 \to 11$: $\rho = 2025/3025 \approx 0.67$, \;\; $\Delta^* \approx 0.49$;
    \item $z=15 \to 16$: $\rho = 11025/14400 \approx 0.77$, \;\; $\Delta^* \approx 0.31$;
    \item $z=20 \to 21$: $\rho = 36100/44100 \approx 0.82$, \;\; $\Delta^* \approx 0.22$.
\end{itemize}
At $z=5 \to 6$, the case used in the empirical validation of
Section 4.1, the threshold $\Delta^*$ is loose: only 44\% of the
$AAC^{(II)}$ value at $z=6$ is already fixed at $z=5$, so that $\Delta^* > 1$ and it is in principle possible
that the extension of design $B$ outperforms the extension of design $A$ at
step $z+1$. From $z \approx 10$ onward the inherited fraction $\rho$
exceeds two thirds and from $z \approx 15$ a gap of more than a
third in $AAC^{(II)}$ at step $z$ virtually guarantees that the ranking is
preserved at step $z+1$. 
\\ \\
Because $GM^{(B)} = \sqrt{AAC^{(B)} \cdot RMS^{(B)}}$ is increasing in both
arguments, a design that leads on both $AAC^{(B)}$ and $RMS^{(B)}$ at step $z$
also leads on $GM^{(B)}$. The partial enumeration procedure, which retains only
the top fraction of designs at each step, is therefore likely to select the
best designs in terms of the $GM^{(B)}$ criterion.
\newpage
\refstepcounter{section}
\section*{Appendix F}
\label{app:ap3}
Let $W_{1}$ be a weighing design of size $m_1 \times z$ with weight $k_1$ and $W_{2}$ a weighing design of size $m_2 \times z$ with weight $k_2$. A larger weighing design of size $(m_1+m_2) \times z$ can be obtained by stacking the two designs: $$W=\begin{bmatrix}
    W_{1} \\W_{2}
\end{bmatrix}.$$
To show that $W$ is a weighing design, consider its two defining properties. Each column of $W_{1}$ contains exactly $k_1$ non-zeros entries and each column of $W_{2}$ contains exactly $k_2$ non-zero entries. Hence, every column of $W$ contains exactly $k_1+k_2$ non-zero entries. 
Moreover, $$W'W=(W_{1})'W_{1}+(W_{2})'W_{2}=I_{k_1}+I_{k_2}=I_{k_1+k_2}.$$
This shows that the columns of $W$ remain orthogonal. $W$ satisfies both defining properties of a weighing design (see definition on page 7) and is therefore itself a valid weighing design.
\newpage
\refstepcounter{section}
\section*{Appendix G}
\label{app:dsd}
This appendix reports the detailed comparison between OMARS designs
and DSDs for the same number of runs
and factors. The comparison complements the discussion in
Section 4.3.2 with the full numerical evidence.
\\ \\
The comparison is carried out at four factor counts ($k = 5,\ 8,\ 9,\ 12$) with
$n = 96$ runs in each case. This choice is motivated by two
considerations. First, $n = 96$ corresponds to the run size used throughout Section 4.3.2. Second, the factor counts cover the
two intended uses of the designs. For $k = 5$ and $k = 8$, the full
second-order model is estimable under both designs, which enables a direct
comparison of the D-, A- and G-efficiencies. For $k = 9$ and $k = 12$, the full second-order model is not estimable for either design at this run size and the comparison for these cases is based on the aliasing measures alone. These two cases correspond to a screening use of the designs, where the number of model parameters is large relative to the run size.
\\ \\
The OMARS designs are selected from the catalog as the best design according
to the criterion of Section 4.3.1 at each $(m=48, z=k)$ combination,
followed by foldover to obtain the $96 \times k$ design. The DSDs are constructed following the conference-matrix framework of \cite{Xiao2012-em}. The construction uses the first $k$ columns of the Paley skew-symmetric conference matrix of order $48$. 
\\ \\
For $k = 5$ and $k = 8$, the D-, A- and G-efficiencies are computed for the
full second-order model. For $k = 9$ and $k = 12$, these efficiencies are not
defined, as explained above. The average absolute correlations within the II,
IQ and QQ blocks are reported for all four factor counts, to expose the source of the differences between the two designs. The GM values of the three blocks,
computed as defined in Section 4.3.1, are also reported.
Table~\ref{tab:dsd-eff} reports the efficiencies,
Table~\ref{tab:dsd-aliasing} the average absolute correlations and
Table~\ref{tab:dsd-gm} the GM values.
\begin{table}[!htbp]
\centering
\caption{D-, A- and G-efficiencies (in \%) for the full second-order model}
\label{tab:dsd-eff}
\small
\begin{tabular*}{\linewidth}{@{\extracolsep{\fill}}cc ccc@{}}
\toprule
$k$ & Design & D-eff & A-eff & G-eff \\
\midrule
5  & DSD   & 36.65 & 3.16 & 40.32 \\
5  & OMARS & 44.68 & 9.27 & 86.39 \\
\midrule
8  & DSD   & 33.38 & 0.58 & 81.60 \\
8  & OMARS & 37.25 & 2.79 & 84.42 \\
\bottomrule
\end{tabular*}
\end{table}
\vspace{-1 cm}
\\
For $k = 5$ and $k = 8$, where the full second-order model is estimable, the
OMARS designs achieve higher D-, A- and G-efficiencies than the corresponding
DSDs. The OMARS designs also achieve lower average II correlations at $k = 5$,
$8$ and $9$, while, at $k = 12$, the average II correlations are essentially
equal ($0.091$ versus $0.089$). The OMARS designs also achieve lower average IQ correlations at
all four factor counts. The DSDs achieve a lower average QQ correlation, which
remains constant at $0.021$ across all factor counts. 
\\ \\
Turning to the GM values, at $k = 5$ and $k = 8$, the OMARS designs have a
higher $GM^{(IQ)}$ than the DSDs despite their lower average IQ correlations.
This is because the GM also involves the RMS values, which is more
sensitive to a small number of large correlations: at $k = 8$, the IQ root mean
square equals $0.155$ for the OMARS design against $0.129$ for the DSD. Under
the GM ranking criterion, which considers the II block first, the OMARS designs
have a clear edge over the DSDs at $k = 5$, $8$ and $9$, while the DSD is
slightly better than the OMARS design at $k = 12$.
\begin{table}[!h]
\centering
\caption{Average absolute correlations within the II, IQ and QQ blocks}
\label{tab:dsd-aliasing}
\small
\begin{tabular*}{\linewidth}{@{\extracolsep{\fill}}cc ccc@{}}
\toprule
$k$ & Design & II Avg & IQ Avg & QQ Avg \\
\midrule
5  & DSD   & 0.038 & 0.089 & 0.021 \\
5  & OMARS & 0.027 & 0.079 & 0.091 \\
\midrule
8  & DSD   & 0.073 & 0.112 & 0.021 \\
8  & OMARS & 0.058 & 0.096 & 0.091 \\
\midrule
9  & DSD   & 0.080 & 0.116 & 0.021 \\
9  & OMARS & 0.053 & 0.085 & 0.091 \\
\midrule
12 & DSD   & 0.089 & 0.124 & 0.021 \\
12 & OMARS & 0.091 & 0.100 & 0.113 \\
\bottomrule
\end{tabular*}
\end{table}
\begin{table}[!h]
\centering
\caption{GM values of the II, IQ and QQ blocks}
\label{tab:dsd-gm}
\small
\begin{tabular*}{\linewidth}{@{\extracolsep{\fill}}cc ccc@{}}
\toprule
$k$ & Design & $GM^{(II)}$ & $GM^{(IQ)}$ & $GM^{(QQ)}$ \\
\midrule
5  & DSD   & 0.050 & 0.102 & 0.021 \\
5  & OMARS & 0.037 & 0.113 & 0.091 \\
\midrule
8  & DSD   & 0.090 & 0.120 & 0.021 \\
8  & OMARS & 0.076 & 0.122 & 0.091 \\
\midrule
9  & DSD   & 0.097 & 0.123 & 0.021 \\
9  & OMARS & 0.075 & 0.114 & 0.091 \\
\midrule
12 & DSD   & 0.106 & 0.130 & 0.021 \\
12 & OMARS & 0.110 & 0.123 & 0.127 \\
\bottomrule
\end{tabular*}
\end{table}
\vspace{-1 cm}
\\
The lower average QQ correlation of the DSDs should not, however, be
interpreted as a more precise estimation of the quadratic effects. In a DSD, each factor takes its middle level in only two of the 96 runs, whereas in the OMARS designs each factor takes its middle level in eight runs.
This difference is reflected in
the standard errors of the estimated effects. 
Table~\ref{tab:dsd-se} reports the average standard error per effect type, in units of $\sigma$, the standard deviation of the error terms in the model, for $k = 5$ and $k = 8$.
\begin{table}[!h]
\centering
\caption{Average standard error per effect type for the full second-order
model, in units of the standard deviation $\sigma$ of the random errors}
\label{tab:dsd-se}
\small
\begin{tabular*}{\linewidth}{@{\extracolsep{\fill}}cc ccc@{}}
\toprule
$k$ & Design & Main & Interaction & Quadratic \\
\midrule
5 & DSD   & 0.103 & 0.112 & 0.786 \\
5 & OMARS & 0.107 & 0.124 & 0.445 \\
\midrule
8 & DSD   & 0.103 & 0.222 & 1.744 \\
8 & OMARS & 0.107 & 0.232 & 0.999 \\
\bottomrule
\end{tabular*}
\end{table}
\vspace{-1 cm}
\\
The quadratic effects in the DSDs are estimated with standard errors approximately $75 \%$ larger than those obtained under the OMARS designs, while the main effects and interactions are estimated with comparable precision across both designs. Consequently, the low average QQ correlation observed for the DSDs is not indicative of accurate quadratic effect estimation but instead reflects quadratic columns that are nearly constant in the model matrix. The greater number of middle-level runs in the OMARS designs provides a substantially higher precision for estimating quadratic effects.
\newpage
\refstepcounter{section}
\section*{Appendix H}
\label{app:ap5}
Consider an OMARS design with $r$ runs and $c$ factors. Under a second-order model, the total number of model parameters $p$ (columns of the model matrix \textbf{X}) is $$p=1+c+\frac{c(c-1)}{2}+c=\frac{c^2+3c+2}{2},$$ corresponding respectively to the intercept, main effects, two-factor interactions and quadratic terms. For all parameters to be estimable, the model matrix must have full column rank, i.e. $$rank(\textbf{X})=p.$$
Because the rank cannot exceed the number of available independent run directions and the foldover structure consumes $f$ fake factor degrees of freedom \citep{Stallrich2025-vc}, the effective estimable rank of an OMARS design is bounded by $$rank(\textbf{X}) \le r-f,$$ where the number of fake factor degrees of freedom is given by $$f=\frac{r}{2}-c.$$
Therefore, a necessary condition for the estimability of all model effects is $$p \le r-(\frac{r}{2}-c)=\frac{r}{2}+c,$$ and thus $$\frac{c^2+3c+2}{2} \le \frac{r}{2}+c.$$ Solving the corresponding quadratic equation yields the admissible range for $c$. The two real roots of equality are $$c=\frac{-1 \pm \sqrt{4r-7}}{2}.$$ 
Since the number of factors must be non-negative, only the upper root is relevant. As a result, the largest number of columns (factors) that allows estimation of all effects in the second-order model is $${c_{max}=\left\lfloor \frac{-1+\sqrt{4r-7}}{2}\right\rfloor}.$$ Respecting $c_{max}$ is a necessary condition for the estimability of the full second-order response surface model. For specific values of $r$ and the corresponding $c_{max}$, see Table \ref{tab:cmax} in Section 4.3.2.
\newpage
\bibliography{sn-bibliography}
\end{document}